\documentclass[prd,twocolumn,preprintnumbers,nofootinbib,superscriptaddress]{revtex4}
\usepackage[latin1]{inputenc}
\usepackage{amsfonts}
\usepackage{subfigure}
\usepackage{hyperref}
\usepackage{bookmark}
\usepackage{slashbox}
\usepackage{dcolumn}
\usepackage{bm}
\usepackage{latexsym}
\usepackage{epsfig}
\usepackage{graphicx,multirow}
\usepackage{units}
\usepackage{color,soul}
\usepackage{amssymb}
\usepackage{amsmath}
\usepackage{enumerate}
\usepackage{textcomp}
\usepackage{mathtools}
\graphicspath{ {figures/} }
\def\re#1{(\ref{#1})}

\newcommand{\be}{\begin{equation}}
\newcommand{\ee}{\end{equation}}
\newcommand{\ba}{\begin{eqnarray}}
\newcommand{\ea}{\end{eqnarray}}

\newcommand{\htildesq}{{\tilde h}^2}

\begin{document}
\preprint{CERN-TH-2017-117}
\title{Failures of homogeneous and isotropic cosmologies  in \\ Extended Quasi-Dilaton Massive Gravity}
 
\author{Stefano Anselmi}
\email{stefano.anselmi@iap.fr}
\affiliation{LUTH, UMR 8102 CNRS, Observatoire de Paris, PSL Research University, Universit\'e Paris Diderot, 92190 Meudon -- France}
\affiliation{Institut d'Astrophysique de Paris, CNRS UMR 7095 and UPMC, 98bis, bd Arago, F-75014 Paris -- France}
\affiliation{Department of Physics/CERCA/Institute for the Science of Origins, Case Western Reserve University, Cleveland, OH 44106-7079 -- USA}

\author{Saurabh Kumar}
\email{saurabh.kumar@case.edu}
\affiliation{Department of Physics/CERCA/Institute for the Science of Origins, Case Western Reserve University, Cleveland, OH 44106-7079 -- USA}

 \author{Diana L\'opez Nacir}\email{diana.laura.lopez.nacir@cern.ch}
 \affiliation{Theoretical Physics Department, 
CERN, CH-1211 Gen\`eve 23, Switzerland}
 
\author{Glenn D. Starkman}
\email{glenn.starkman@case.edu}
\affiliation{Department of Physics/CERCA/Institute for the Science of Origins, Case Western Reserve University, Cleveland, OH 44106-7079 -- USA}

\date{\today}

\begin{abstract}
We analyze the Extended Quasi-Dilaton Massive Gravity model around a Friedmann-Lema\^\i tre-Robertson-Walker cosmological background. We present a careful  stability analysis of asymptotic fixed points.  We find that the traditional fixed point cannot be approached dynamically,
except from a perfectly fine-tuned initial condition involving both the quasi-dilaton
and the Hubble parameter. 
A less-well examined fixed-point solution, where the time derivative of the 0-th St\"uckelberg 
field vanishes $\dot\phi^0=0$, encounters no such difficulty,
and the fixed point is an attractor in some finite region of initial conditions. We examine the question of the presence of a Boulware-Deser ghost in the theory.
We show that the additional constraint which generically allows for the elimination of the Boulware-Deser mode 
is    {\it only} present under special initial conditions.  
We find that the only possibility  corresponds  to the traditional fixed point and the  initial conditions are  the same
fine-tuned conditions that allow the fixed point to be approached dynamically. 

\end{abstract}
\maketitle
\section{Introduction}

In the standard cosmological model, 
the accelerated expansion of the universe is attributed 
to the cosmological constant $\Lambda$. 
However, to match the observed expansion, 
$\Lambda$ must be of the order of $10^{-122}$ in Planck units, 
which raises a fine-tuning problem. 
A possible alternative is to modify general relativity (GR) at large distances or low momenta. 
A massive spin-2 field theory, known as the dRGT theory~\cite{dRGT,HasanRosen}, 
is a theoretically well-motivated modification of GR. 
However, the dRGT theory does not admit a flat Friedmann-Lema\^\i tre-Robertson-Walker (FLRW) solution 
with an expanding scale factor~\cite{adoteq0}. 
A modification to dRGT gravity, known as Quasi-Dilaton Massive Gravity (QDMG), 
was proposed in~\cite{damico} and provides homogeneous and isotropic expanding solutions. 
It was later shown~\cite{Anselmi:2015zva} 
that the parameters of QDMG have to be finely tuned 
in order to match the observed expansion history of the universe. 
More disastrously, the results in~\cite{damico2,emir} indicate 
that the scalar perturbations in QDMG 
acquire a wrong sign kinetic term at short scales. 
A further modification, Extended QDMG (EQDMG), was proposed in~\cite{AdF-SM}
and has scalar perturbations that are thought to be stable at all momentum scales. 
The standard fixed-point cosmological solution of EQDMG has a  de Sitter metric,
and thus appears to be a good candidate for late-time cosmology.

  EQDMG  only differs from QDMG by the addition to the action of one operator involving the Quasi-Dilaton  (QD) field and a new free parameter  $\alpha_{\sigma}$. Naively, in the limit $\alpha_{\sigma}\to 0$, EQDMG reduces to QDMG, but actually the limit is very subtle. Indeed, there are controversial results in the literature regarding whether or not EQDMG contains  an unavoidable  additional degree  of freedom, which would correspond to the  Boulware-Deser ghost (BD) \cite{Kluson:2013jea,Mukohyama:2013raa,Heisenberg:2015voa}.

In this paper, for the first time, we:
\begin{enumerate}
    \item assess the stability of the standard fixed-point solutions 
    (referred to as Case 1 in this paper) and show that this assessment requires a non-standard approach;
    \item demonstrate that the Case 1 fixed points cannot be approached dynamically, 
    due to an unavoidable singularity in the dynamical equations;
    \item perform a comprehensive study of a new branch of solutions (referred to as Case 2), first proposed by~\cite{Gumrukcuoglu:2016hic} but largely ignored in the literature, 
    and show that it provides stable and dynamically attainable fixed-point solutions;
 \item  show for a flat FLRW universe, the fact that the background equations are satisfied does not guarantee the presence of the additional constraint necessary to eliminate the BD mode (in agreement with the results of \cite{Kluson:2013jea}, but in disagreement with the  computations in \cite{Heisenberg:2015voa});
 \item  find that the only branch of solutions for which the additional constraint exists corresponds to Case 1;
 \item argue that, in order to avoid a BD ghost, the initial values of certain EQDMG dynamical variables must be extremely fine-tuned;
 \item verify that the same fine-tuned initial conditions also allow the fixed point to be approached dynamically.
\end{enumerate}  

The paper is organized as follows. 
In Section~\ref{Formalism} and~\ref{Background}, 
we summarize the theory of EQDMG. 

In Section~\ref{DynEqnsVar}, 
we define the dynamical variables and provide the relevant background equations. 

In Section~\ref{fixedpoint}, 
we find the fixed-point and de Sitter solutions of the dynamical equations, 
and show that they are equivalent to one another (provided the Hubble rate is positive). 
We identify four independent fixed-point cases, 
each of which is studied in greater detail in the sections that follow. 
 
In Section~\ref{stability}, 
we introduce the standard procedure for analyzing the stability of the fixed-point solutions
for the background. 
We discuss the inadequacy of this procedure for the Case 1 fixed points 
and provide an augmented framework.

In Section~\ref{scanning}, we present the results of our numerical search for viable
parameters for the EQDMG theory. We find that except for a very specific, precisely fine-tuned
initial displacement away from the fixed-point values of Case 1 (explained in Section \ref{bdghost}), the fixed points cannot be reached in the asymptotic future.

In Section~\ref{bdghost}, 
we further study the perturbative stability of the scalar sector of the theory, 
both in the vacuum case and with matter. 
We identify the conditions on the dynamical variables
required to avoid the BD ghost,
and show that the fact the background equations are satisfied does not guarantee the validity of the conditions.
 Cases 1 turns out to be  the only case for which the additional constraint necessary to eliminate the BD mode can be obtained  by   an  appropriate choice of   initial conditions. 
However, those conditions appear to represent a difficult fine-tuning of all the degrees of freedom. 

In Section~\ref{cnclsns} we present our conclusions. We provide some detailed calculations and consider the special case of Minkowski solutions in the appendices.
\section{Formalism}
\label{Formalism}
We consider the action for the  extended quasi-dilaton theory \cite{AdF-SM}: 
\begin{eqnarray}
S & = & S_{EH}+S_{m}+S_{\sigma}
= \frac{M_{{\rm Pl}}^{2}}{2}\int d^{4}x\sqrt{-g}\biggl[R\\ \nonumber
&-&\frac{\omega}{M_{{\rm Pl}}^{2}}\partial_{\mu}\sigma\partial^{\mu}\sigma+2m_{g}^{2}(\mathcal{L}_{2}+\alpha_{3}\mathcal{L}_{3}+\alpha_{4}\mathcal{L}_{4})\biggr]+S_{m},\label{eq:action}
\end{eqnarray}
where $M_{\rm Pl}$ is the Planck mass and,  
in addition to the Einstein Hilbert action $S_{EH}$, 
we have the contribution $S_m$ of the matter sector,  
and a quasi-dilaton contribution $S_{\sigma}$.
Here
\begin{eqnarray}
\mathcal{L}_{2} & \equiv & \frac{1}{2}\,([\mathcal{K}]^{2}-[\mathcal{K}^{2}])\,,\\
\mathcal{L}_{3} & \equiv & \frac{1}{6}\,([\mathcal{K}]^{3}-3[\mathcal{K}][\mathcal{K}^{2}]+2[\mathcal{K}^{3}])\,,\\
\mathcal{L}_{4} & \equiv & \frac{1}{24}\,([\mathcal{K}]^{4}-6[\mathcal{K}]^{2}[\mathcal{K}^{2}]+3[\mathcal{K}^{2}]^{2}\\ \nonumber
&+&8[\mathcal{K}][\mathcal{K}^{3}]-6[\mathcal{K}^{4}])\,,
\end{eqnarray}
 with  square brackets denoting a trace, and
\begin{eqnarray}
\label{ktensor}
\mathcal{K}_{\ \nu}^{\mu}&=&\delta_{\ \nu}^{\mu}-e^{\sigma/M_{{\rm Pl}}}\left(\sqrt{g^{-1}{f}}\right)_{\ \ \nu}^{\mu}\,,\\
{f}_{\mu\nu}&\equiv&\eta_{ab}\partial_{\mu}\phi^{a}\partial_{\nu}\phi^{b}-\frac{\alpha_{\sigma}}{M_{{\rm Pl}}^{2}m_{g}^{2}}e^{-2\sigma/M_{{\rm Pl}}}\partial_{\mu}\sigma\partial_{\nu}\sigma.
\end{eqnarray}  
$S_{\sigma}$ includes five new fields:
$\sigma$ is the quasi-dilaton  scalar field
and $\phi^{a}$ ($a=0,\cdots,3$) are the four St\"uckelberg fields. 
It also depends on  the coupling constants 
$\alpha_{\sigma}$, $\alpha_{2}$ and $\alpha_{3}$, 
and on the graviton mass $m_{g}$. 
For $\alpha_{\sigma}=0$ one recovers the standard quasi-dilaton theory.
 
In the space of
St\"uckelberg fields, the theory  enjoys the Poincare symmetry
\begin{equation}
\phi^{a}\to\phi^{a}+c^{a}\,,\qquad\phi^{a}\to\Lambda_{b}^{a}\phi^{b}\,,
\end{equation} and a  global symmetry given by
\begin{equation}
\sigma\to\sigma+\sigma_{0}\,,\qquad\phi^{a}\to e^{-\sigma_{0}/M_{{\rm Pl}}}\,\phi^{a}\,,\label{eqn:symmetry}
\end{equation}  with $\sigma_{0}$  an arbitrary constant.

\section{Background}
\label{Background}
We  consider  a spatially flat 
Friedmann-Lema\^\i tre-Robertson-Walker (FLRW) ansatz, for which
\begin{eqnarray}
ds^{2} & = & -\mathcal{N}(t)^{2}dt^{2}+a(t)^{2}\delta_{ij}dx^{i}dx^{j}\,,\\
\phi^{0} & = & \phi^{0}(t)\,,\\
\phi^{i} & = & x^{i}\,,\\
\sigma & = & \bar{\sigma}(t)\,.
\end{eqnarray}
The fiducial metric $f_{\mu\nu}$ reduces to 
\begin{equation}
{f}_{00}=-n(t)^2\,, \quad {f}_{ij}=\delta_{ij}\,,
\end{equation}
where
\begin{equation}
\label{defn}
 n(t)^2 \equiv \bigl(\dot{\phi}^{0}\bigr)^{2}
 +\frac{\alpha_{\sigma}}{M_{\rm Pl}^2m_g^2}\,e^{-2\bar{\sigma}/M_{\rm Pl}}
 {\dot{\bar{\sigma}}}^2\,.
\end{equation}
The minisuperspace action for the background can be written as
\begin{eqnarray}
S/V & = & M_{{\rm Pl}}^{2}\int dt\biggl[-3 \frac{a^3}{\mathcal{N}}\left(\frac{\dot{a}}{a}\right)^2+a^3\frac{w}{M_{\rm Pl}}\frac{\dot{\sigma}^2}{2\mathcal{N}}\\ \nonumber
& + &\mathcal{N} a^3 m_{g}^{2}(\mathcal{L}_{2}+\alpha_{3}\mathcal{L}_{3}+\alpha_{4}\mathcal{L}_{4})\biggr],
\label{eq:actionB}
\end{eqnarray}
where
\begin{eqnarray}
\mathcal{L}_{2} &= &3(X-1)(-2+X(1+r))\,,\\
\mathcal{L}_{3} & =&-(X-1)^2(-4+X(1+3r))\,,\\
\mathcal{L}_{4} & = &(X-1)^3(-1+rX)\,,
\end{eqnarray}
and we have defined
\begin{eqnarray}
\label{xsigma}
X & \equiv & \frac{e^{\bar{\sigma}/M_{\rm Pl}}}{a}\,,\\
r & \equiv & \frac{n}{\mathcal{N}}\, a\,\label{rna}.
\end{eqnarray}
It is worth pointing out here that in~\re{ktensor},  
\[
\left(\sqrt{g^{-1}f}\right)^{\mu}_{\nu}=
  \begin{bmatrix}
    \pm{\nicefrac{n}{\mathcal{N}}} & 0 & 0 & 0 \\
    0 & \pm{\nicefrac{1}{a}} & 0 & 0\\
    0 & 0 & \pm{\nicefrac{1}{a}} & 0 \\
    0 & 0 & 0 & \pm{\nicefrac{1}{a}}
  \end{bmatrix}
\]
and we make the $(+,+,+,+)$ choice following~\cite{2015PhRvD..91l1502C}. This gives us
\begin{equation}
r>0.
\end{equation}
 $r=0$ represents a determinant singularity
in either or both $f_{\mu\nu}$ (if $n=0$) or $g_{\mu\nu}$ (if $a=0$) -- a spacelike hypersurface 
where the dimensionality  of the metric changes.  
The stability of the theory across that hypersurface is unclear 
\cite{2015PhRvD..92d4024M}.  Indeed, we find that when $r$ approaches too close to zero, 
our numerical integrations of the dynamical equations become unstable;
the instability in the numerical noise may be due to an underlying instability in the theory. 
We insist here that $r\neq0$.  

\section{Dynamical Equations and Variables}
\label{DynEqnsVar}

We next set out the dynamical variables describing the background,
and the dynamical equations describing their time evolution.

Varying the action with respect to $\phi^0(t)$ leads to the constraint equation
\begin{equation}\label{Constraint}
    \partial_t\left[ \frac{\dot{\phi^0}}{n}a^4\, G_2(X)\right] = 0\,, 
\end{equation}
where  $G_2(X)=X(1-X)J(X)$, with
\begin{equation}
\label{jterm}
    J(X) \equiv 3+3(1-X)\alpha_{3}+(1-X)^{2}\alpha_{4}\,.
\end{equation}
This suggests that it will be useful to introduce as one of the dynamical
variables 
\begin{equation}
\label{ydef}
    y(t)\equiv\frac{\dot{\phi^0}(t) }{n}G_{2}(X) \,.
\end{equation}
The solution of (\ref{Constraint}) is immediately 
\begin{eqnarray}\label{yc}
    y&=&\frac{C}{a^4}\,.
\end{eqnarray}

We see that, in any reasonable cosmological context, $y\to0$ in the asymptotic  future,
and that this can be achieved by one (or more) of four quantities approaching or equalling zero:
$J(X(t))$, $\dot{\phi^0}(t)$, $X(t)$, or $X(t)-1$.
These four cases will drive our analysis.

Anticipating that it will be convenient to regard $X$ as a dynamical variable,
we differentiate~\re{xsigma} with respect to time to get
\begin{equation}
\label{sigdot}
\dot{X}=X\left(\frac{\dot{\sigma}}{M_{Pl}}-H\right)\,.
\end{equation}
$H(t)\equiv{\dot a}/a$ is the Hubble parameter.

Varying the action with respect to the lapse $\mathcal{N}(t)$ 
and using time reparametrization invariance to set $\mathcal{N}(t)=1$,
we obtain the Friedmann equation:
\begin{equation}\label{Friedman}
    3H^2=\frac{\omega}{2}\left(\frac{\dot\sigma}{M_{\rm Pl}}\right)^2
    + 3m_g^2  G_1(X)+\frac{\rho_m}{M_{\rm Pl}^2}+\frac{\rho_r}{M_{\rm Pl}^2}\,,
\end{equation} where  
\begin{equation}
\label{G1}
    G_1(X)\equiv\frac{1}{3} (X- 1) \left(J(X) + (X - 1) (\alpha_3 (X - 1) - 3)\right)\,.
\end{equation}

The form of the Friedman equation suggests regarding 
the first two terms on the right-hand side (RHS) of (\ref{Friedman})
as the dark energy density (divided by $M_{\rm Pl}^2$).
For future convenience, 
we represent them separately as  
\begin{equation}
\label{omegaL}
    \Omega_{\Lambda}\equiv\frac{m_g^2}{H^2}G_1(X)\\
\end{equation}
and
\begin{equation}
\label{omegaS}
    \Omega_{\sigma}\equiv\frac{\omega}{6H^2}\left(\frac{\dot{\sigma}}{M_{\rm Pl}}\right)^2\,,
\end{equation}
and define $\Omega_{DE}\equiv \Omega_{\Lambda}+\Omega_{\sigma}$. 

From the conservation of the stress-energy tensor of matter and of radiation we get,
\begin{eqnarray}
    \dot{\rho_m}&=&-3H{\rho_m},\label{omegam}\\
    \dot{\rho_r}&=&-4H{\rho_r}\label{omegar}\,.
\end{eqnarray}
From the conservation of the stress-energy tensor\footnote{
    Note that, using the constraint equation~\re{Constraint}, 
    one can show that the equation obtained by taking 
    the variation of $S_\sigma$ with respect to $\sigma$ is not an independent equation.} 
obtained from $S_{\sigma}$, we get
\begin{equation}\label{conservSigma}
    (\ddot{\sigma}+3H\dot{\sigma})\omega\dot{\sigma}
    +3 M_{\rm Pl} m_g^2(\dot{\sigma}-r H M_{\rm Pl})X G_1'(X)=0,
\end{equation} 
where a prime here stands for a derivative with respect to the argument $X$ of the function.

We select $y,X,\tilde{\Omega}_i=\Omega_i \htildesq (i=DE,m,r)$ 
as our dynamical variables,
where ${\tilde h}\equiv H/m_g$ is the dimensionless Hubble parameter.
We can now express the background evolution equations in terms of 
 $N\equiv log(a)$ 
(giving us $dN=H dt$ assuming $H \neq 0$)\footnote{
The special case $H=0$, i.e. Minkowski space, is discussed in appendix \ref{Minkowski}.
}:
\begin{itemize}
    \item 
    From (\ref{Constraint}), we have immediately
    \be
        \frac{dy}{dN}=-4y \label{yN}\,.
    \ee
    \item The equations for $\tilde{\Omega}_{m}$ and $\tilde{\Omega}_{r}$ are similarly easily obtained from ~\re{omegam} and ~\re{omegar}:
    \begin{eqnarray}
        \frac{ d\tilde{\Omega}_{m}}{dN}&=&-3 \tilde{\Omega}_{m},\label{matter}\\
        \frac{ d\tilde{\Omega}_{r}}{dN}&=&-4\tilde{\Omega}_{r}\label{radiation}\,.
    \end{eqnarray}
    \item Equation~\re{sigdot} can be rewritten
    using~\re{omegaL},~\re{omegaS} and the definition of $\tilde{\Omega}_{DE}$:
    \be
        \frac{d X}{dN}=
        X\left(\pm\sqrt{\frac{6\left(\tilde{\Omega}_{DE}-G_{1}(X)\right)}{\htildesq\omega}}
        -1\right)\label{txx}\,.
    \ee
    The $\pm$ represents the possibility that $\dot\sigma$ can be positive or negative\footnote{
        We will focus our attention below on the positive sign, because the negative sign 
        leads to only an $X=0$ fixed point.
        }.    
    \item Equation~\re{conservSigma} can be rewritten 
    using ~\re{omegaL} and~\re{omegaS}:
     \be
        \label{tildeOmsigmaeqs}
        \frac{ d\tilde{\Omega}_{DE}}{dN}
        =-6\left(\tilde{\Omega}_{DE}-G_{1}(X)\right) 
        +X G_1'(X) \left(r-1\right)\,.
    \ee
\end{itemize}
In the above set of equations one must replace
\begin{itemize}
    \item $\htildesq$ by the Friedmann equation, which now takes the simple form
    \begin{eqnarray}\label{hsqnew}
        \htildesq&=&\tilde{\Omega}_{DE}+\tilde{\Omega}_{m}+\tilde{\Omega}_{r}\,,
    \end{eqnarray} 
    \item and (combining \re{defn}, \re{rna} and \re{yc}) $r$ with\footnote{
        For $\alpha_{\sigma}=0$,
        $r$ cannot be determined from ~\re{rdef},
         because~\re{ydef} gets reduced to $y=G_{2}(X)$. 
         Thus, we can no longer use~\re{tildeOmsigmaeqs}, 
         and the above system of evolution equations is not well-equipped to handle this case. 
         In fact, this limit gives us the Quasi-Dilaton theory and the evolution of the dynamical variables have been previously studied by~\cite{Anselmi:2015zva} and \cite{2013PhRvD..87l3536G}.} 
    \begin{eqnarray}
    \label{rdef}
    r&=&+\sqrt{\frac{6\left(\tilde{\Omega}_{DE}-G_{1}(X)\right)\alpha_{\sigma}}{\omega X^{2}\left(1-\left(\frac{y}{G_2(X)}\right)^2\right)}}.
    \end{eqnarray}
\end{itemize}

The argument of the square root on the right-hand side of ~\re{rdef} 
must be positive for $r$ to be real.
{\em The reality of $r$ is a condition on the dynamical variables 
that must be checked, 
in case (as we find below) it is not automatically satisfied. }
In particular, we see that problems may arise if $\left(y/G_2(X)\right)^2\to1$.
\section{Fixed-point Analysis}
\label{fixedpoint}

In this section, we evaluate the dynamical variables 
when their $N-$derivatives vanish in~\re{yN}-\re{tildeOmsigmaeqs}. 
We term the values of the dynamical variables in this limit as fixed points. 

In the fixed-point limit, the left-hand sides of equations~\re{yN}-\re{radiation} vanish,
giving us $y_{FP}=0$, and also $\tilde{\Omega}_{m,FP}=\tilde{\Omega}_{r,FP}=0$. 
From~\re{hsqnew}, we learn that $\tilde{\Omega}_{DE,FP}=\tilde{h}_{FP}^2$. 

The solutions to~\re{yN}-\re{radiation} are
\begin{eqnarray}
y=y_0e^{-4N},\hskip 2pt
\tilde{\Omega}_{m}=\tilde{\Omega}_{m0}e^{-3N},
\hskip 2pt\tilde{\Omega}_{r}=\tilde{\Omega}_{r0}e^{-4N},
\end{eqnarray}
where $y_0$, $\tilde{\Omega}_{m0}$ and $\tilde{\Omega}_{r0}$ 
are the corresponding initial values. 
Thus fixed points occur in the asymptotic future, i.e. as $N\to\infty$ and so $a \to \infty$. 

The left-hand side of ~\re{txx} 
vanishes at the fixed point, implying that $X_{FP}$ is a constant.
If $X_{FP}\neq0$, the right-hand side of ~\re{txx} 
(and if $X_{FP}=0$, then the right-hand side of~\re{tildeOmsigmaeqs})
provides us with
\be
    \label{hsqfp}
    \tilde{\Omega}_{DE,FP}=\tilde{h}_{FP}^2 = \begin{cases}
        \frac{G_{1}(X_{FP})}{\left(1-\frac{\omega}{6}\right)}, & X_{FP}\neq0\,,\\
        G_1(0), & X_{FP}=0\,,
        \end{cases}
\ee
at the fixed point.
In arriving at~\re{hsqfp}, we take the \textsc{\char13}$+$\textsc{\char13} sign in~\re{txx},
since the \textsc{\char13}$-$\textsc{\char13} sign leads to $X=0$ as the only fixed-point solution.

Notice that, as for getting background fixed-point solutions, both  $0<\omega<6$ and
$\omega\geq6$  are in principle suitable regions in the parameter space,
since $G_{1}(X_{FP})$ can be either positive or negative.
A special value is $\omega=6$, 
in which case ~\re{txx} and~\re{hsqfp}
demand  $G_{1}(X_{FP})=0$.  

Observing that $\htildesq$ is also a constant at the fixed point,
we conclude that  \textit{the fixed points of the evolution equations are de Sitter}. 

We find that the converse is also true:
\textit{the de Sitter solutions of the evolution equations 
are fixed points as we approach the asymptotic future.} 
To prove this, we require that the dynamical variables 
attain the following de Sitter values in the future, 
\begin{eqnarray}
\left(\tilde{\Omega}_{DE},\tilde{\Omega}_m,\tilde{\Omega}_r\right)
&=&\left(\tilde{h}_{FP}^2,0,0\right)\label{fpvals},
\end{eqnarray}
where $\tilde{h}_{FP}^2$ is a constant different from zero. 
In this situation, the left-hand sides of~\re{tildeOmsigmaeqs},~\re{matter} and~\re{radiation} become zero, meaning they are fixed points.

From~\re{yc}, we learn that $y = 0$ in the asymptotic future, 
which means that the left-hand side  of~\re{yN} is also zero. 
The only point left to establish is that $X$ approaches a constant in the future.

From the definitions~\re{defn},\re{rna} and \re{ydef}, 
one can split the fixed-point solution into four cases:
\begin{itemize}
\item Case 1 (the standard case):
    \be
    \label{case1 }
    J(X_{FP})=0\,,
    \ee
    and hence the fixed-point solutions are
\begin{equation}
\label{xpm}
 X_{FP}=X_{\pm} = 1+\frac{3}{2}\frac{\alpha_3}{\alpha_4}\pm{\sqrt{\frac{9\alpha_{3}^{2}}{4\alpha_{4}^2}-\frac{3}{\alpha_4}}}.
 \end{equation}

\item 
    Case 2:
    \begin{equation}
        \label{case2 }
        \left[\frac{\dot{\phi}^0}{n}\right]_{FP} = 
        \sqrt{1-\alpha_{\sigma}\frac{6\left(\tilde{\Omega}_{DE,FP}-G_{1}(X_{FP})\right)}{\omega r^2_{FP} X^2_{FP}}}=0 \,.
    \end{equation}
    
Since the left-hand side  of~\re{tildeOmsigmaeqs} vanishes,~\re{rdef} provides us with the following equation in $X_{FP}$ 
\begin{eqnarray}
    \label{7th}
    &6(\tilde{h}^2_{FP}-G_{1}(X_{FP}))+X_{FP}G'_{1}(X_{FP})\\ \nonumber
    &=G'_{1}(X_{FP})\sqrt{\frac{6}{\omega }\alpha_{\sigma}
    \left(\tilde{h}^2_{FP}-G_{1}(X_{FP})\right)}.
\end{eqnarray}
Squaring~\re{7th} gives us a polynomial equation for $X_{FP}$. 
$X_{FP}$ can be any of the roots of that polynomial.
\item
    Case 3:
    \be
    \label{case3}
    X_{FP}=0\,.
    \ee
\item
    Case 4:
    \be
    \label{case4}
    X_{FP}=1\,.
    \ee
\end{itemize}

For all  cases,
$X$ approaches a constant and thus the left-hand side  of~\re{txx} vanishes in the asymptotic future proving it is a fixed point.

We  analyze the fixed-point solutions in more detail below.

\subsection{Case 1: $J(X_{FP})=0$}

As will become clear below, this case is very subtle. 
Note that these fixed points are the same as ones analyzed  in 
~\cite{Anselmi:2015zva},~\cite{2013PhRvD..87l3536G}   
for the QD theory (i.e., the EQD with $\alpha_{\sigma}=0$).

Requiring that $X_{\pm}$ be real, means\footnote{A special case occurs when 
    $\alpha_{4}=\frac{2\alpha_{3}^{2}}{3}$,
    in which case $G_{1}(X)=0=J(X)$. We are left with
    \be
    \label{g1g20}
    \left(X_{FP},r,_{FP},\omega\right) = \left(1+\frac{3}{{\alpha}_{3}},1+\frac{6\htildesq}{X_{FP}^{2}},6\right).
    \ee
    This solution for $\omega=6$ and $\htildesq$ is indeterminate because of simplifications occurring in \re{txx}. 
Therefore, on the $\alpha_{4}=\frac{2\alpha_{3}^{2}}{3}$ hypersurface in the parameter space, the dynamical equations lose predictive power.}
\begin{eqnarray}
\alpha_{4}\leq\frac{3\alpha_{3}^{2}}{4}\,.
\end{eqnarray}

From~\re{tildeOmsigmaeqs} and~\re{hsqfp}, 
we  get the same expression as in the QD theory 
for the fixed-point limit of $r$ (assuming $\omega\neq6$):
\be
r_{FP}=1+\frac{\omega G_{1}(X_{FP})}{X_{FP}G'_{1}(X_{FP})(1-\nicefrac{\omega}{6})}.
\label{rfp}
\ee
Note that this  expression is valid  for Cases 1, 2 and 4 (not $X=0$), 
provided $\omega\neq6$ and $G'_{1}(X)\neq0$.

Unlike QD  theory,  
since  $\alpha_{\sigma}\neq 0$ 
we can use~\re{ydef} to obtain~\re{rdef}, 
which gives  $r$ in terms of the dynamical variables. 
If the system is to evolve  towards its expected fixed point,  
$r$ must approach $r_{FP}$.  
Therefore,
 at the fixed point  $z^2\equiv\left(\frac{y}{G_{2}(X)}\right)^2$ 
 should approach   
\begin{eqnarray}
\label{yg2}
z_{FP}^2&\equiv&\left(\frac{y_{FP}}{G_{2}(X_{FP})}\right)^2\\
&=&1-\alpha_{\sigma}\frac{G_{1}(X_{FP})(1-\nicefrac{\omega}{6})(G'_{1}(X_{FP}))^{2}}{\left[X_{FP}G'_{1}(X_{FP})(1-\nicefrac{\omega}{6})+\omega G_{1}(X_{FP})\right]^2}\,.\nonumber
\end{eqnarray}

In order for this approach to be smooth, one also needs
\be
\frac{dz}{dN}=-\left[4+\frac{G_2'(X)}{G_{2}(X)}\frac{dX}{dN}\right]z\to0.
\ee  
One possibility is that $z\to0$,
which requires a fine tuned relation among the parameters of the model.
Since this is subsumed in Case 2 anyway,  we will not analyze this particular case any further. 
A second possibility is that the quantity in square brackets 
approaches zero  in the fixed-point limit.
This implies  a constraint  equation for the dynamical variables near the fixed point.
We note that in previous  literature this constraint equation has been assumed to hold  
for the full dynamics  with no justification \footnote{
    This constraint is eq.~(7) of \cite{Kahniashvili:2014wua}.}.  
In  such a case, the evolution of the dynamical system near the fixed point 
could be described in terms of 4 instead of 5 dynamical variables;
i.e., near the fixed-point limit  the evolution  would be driven by the same dynamical equations as in the QD theory.  
However,  {\it a priori} there is no reason to expect this condition to be valid
 for $\alpha_{\sigma}\neq0$.

\subsection{Case 2: $\left[\frac{\dot{\phi}^0}{n}\right]_{FP} =0$}

For Case 2, one has to solve~\re{case2 } to get the fixed-point value of $X$.
We must take care  that, after solving for $X_{FP}$, 
the sign of $6(\tilde{h}^2_{FP}-G_{1}(X_{FP}))+X_{FP}G'_{1}(X_{FP})$ 
should be the same as the sign of $G'_{1}(X_{FP})$\footnote{
    There is a special solution of~\re{7th} where $G_{1}(X)=G'_{1}(X)=0$. 
    This is possible only when
   $3+2\alpha_{3}+3\alpha_{3}^{2}-4\alpha_{4} = 0$
    and the common  root of $G_{1}(X)$ and $G'_{1}(X)$ is   $ X_{FP}=(3+5\alpha_{3}+2\alpha_{4})/\left(2\alpha_{3}+2\alpha_{4}\right)$, which corresponds to $\tilde{h}_{FP}=0.$
        Thus we are returned to the special case $h=0$ discussed in Appendix \ref{Minkowski}.
     }.

It is worthwhile noting that for Case 2, $\alpha_\sigma>0$. 
(We omit the QD case, $\alpha_\sigma=0.$)
This can be seen by inspection of equation~\re{case2 },
recalling that 
\be
\tilde{\Omega}_{DE}-G_1(X)=\frac{\omega}{6}\left(\frac{\dot\sigma}{m_gM_{\rm Pl}}\right)^2\geq0\,.
\nonumber
\ee

\subsection{Case 3: $X_{FP}=0$}

From \re{hsqfp} and the definition of $\tilde{\Omega}_{DE}$,~\re{omegaL} and~\re{omegaS}, we get 
\be
\label{sigdotx0}
\dot{\sigma}_{FP}=0\,.
\ee 
We find that $r_{FP}$ is indeterminate from both~\re{tildeOmsigmaeqs} and~\re{rdef}.

Examining~\re{xsigma}, we see that there are two ways to get $X_{FP}=0$.
The first possibility is that $\sigma\to-\infty$,
in which case we can draw no conclusion from~\re{defn} about
the value of  $\frac{\dot{\phi}^{0}}{n}$ 
at the fixed point.
The second possibility is that 
$n^{2}=(\dot{\phi}^{0})^{2}$ at the fixed point.

\subsection{Case 4: $X_{FP}=1$}

Since $G_{1}(1)=0$ and $G_{1}'(1)=1$, 
and since $\Omega_{m}=\Omega_{r}=0$ at the fixed point, 
from~\re{txx} and ~\re{hsqnew}, 
we find that $X_{FP}=1$ requires $\omega=6$. 
Substituting $X=X_{FP}=1$ in~\re{tildeOmsigmaeqs}, we get
\be
r_{FP}=1+6\tilde{h}_{FP}^{2}  
\label{rx1}
\ee
where $\tilde{h}_{FP}$ is indeterminate. Therefore the theory loses its predictive power. For this reason we will not consider this case anymore in the following analysis.

\section{Fixed-Point  Linear Stability}
\label{stability}
We wish to check whether or not 
the fixed-point solutions are attractors in the asymptotic future. 
This would be the case if any small perturbation around the fixed point 
decays to zero asymptotically.

We start with the prescription given by~\cite{Copeland} to evaluate the fixed-point stability.

Let $\mathbf{V}=\left[{y},{X},{\tilde{\Omega}_{DE}},{\tilde{\Omega}_m},{\tilde{\Omega}_r}\right]^{T}$ denote the dynamical variables and $\mathbf{f}(\mathbf{V})$ be the RHS of the first order differential equations. Thus we can express equations~\re{yN}-\re{tildeOmsigmaeqs} as 
\be
\frac{d\mathbf{V}}{dN}=\mathbf{f}(\mathbf{V})\, .
\ee
Assuming small perturbations $\delta{\mathbf{V}}$ around any point, $\mathbf{V}_{0}$, a Taylor expansion of the functions $\mathbf{f}(\mathbf{V})$ gives us
\begin{equation}
\label{taylor}
\frac{d}{dN} \mathbf{\delta V}={\rm \bf M}  \mathbf{\delta V} + \mathbf{f}\left(\mathbf{V}_{0}\right)\, ,
\end{equation}
where ${\bf M}$ is the stability matrix. Its elements are given by
\begin{eqnarray}
\label{stabmat}
M_{ij}&=&\left[\frac{\partial f_{i}(\mathbf{V})}{\partial V_{j}}\right]_{\mathbf{V}=\mathbf{V_0}}, i,j=1,\dots,5
\end{eqnarray}
In Appendix~\ref{StandardStability}, we provide the analytical expressions for the elements of  ${\rm \bf M}$  (Equations~\re{MM22}-\re{MM33}). 

If $\mathbf{V}_{0}$ is a fixed point, then the second term of RHS of~\re{taylor} would vanish. Using the eigenvectors of  ${\rm \bf M}$, one can then find matrix  ${\rm \bf P}$ such that
\begin{eqnarray}
{\rm\bf\mathcal{D}}&=&{\rm\bf P}^{-1}{\rm\bf M}{\rm\bf P},\label{diagonal}\\
&=&\text{Diag}(\lambda_{1},\dots,\lambda_{5})
\end{eqnarray}
where $(\lambda_{1},\dots,\lambda_{5})$ are the eigenvalues given by~\re{eigenvalues}. We define $\delta \mathbf{\bar{V}}$ as
\begin{eqnarray}
&{\rm\bf P} \delta \mathbf{\bar{V}}\equiv\delta\mathbf{{V}},\label{PdelV}
\end{eqnarray}
Thus the solution to~\re{taylor} in the new basis would be 
\be
\delta{\bar{V}_{i}}=e^{\lambda_{i}N}C_{i}.
\ee
where $C_{i}$'s are integration constants. Multiplying the above equation by ${\bf P}$  and thereby returning to the original basis, we get
\be
\delta{V}_{i}=\sum_{j=1}^{5}P_{ij}e^{\lambda_{j}N}C_{j}.
\label{pert_soln}
\ee
Then, for the fixed points to be stable, we require $\delta{\mathbf{V}}$ to approach zero as $N\to\infty$. It can be seen from~\re{pert_soln} that if the eigenvalues of ${\rm\bf M}$ are either real and negative or imaginary with negative real part, the fixed points will be stable or form a stable spiral respectively.   
We find in Appendix~\ref{StandardStability}, that to obtain attractor solutions, $\lambda_{4}$ and $\lambda_{5}$ must be real and negative or complex with negative real parts.
This requires the elements of ${\rm\bf M}$ to satisfy the condition
\be
\label{stable}
0\leq(3+2M_{22})^2+4M_{23}M_{32}<9
\ee
for stable solutions or
\be
\label{stablespiral}
(3+2M_{22})^2+4M_{23}M_{32}<0
\ee
for stable spiral solutions.

\subsection{Case 1: $J(X_{FP})=0$}
\label{case1 stability}
The above-described standard method will not suffice to evaluate fixed-point stability in all cases.
In particular, for $J(X_{FP})=0$
the stability matrix ${\rm\bf M}$ has divergent terms at the fixed point
(see Appendix~\ref{StandardStability} for the exact expressions of its elements). 
This makes the evaluation of the matrix  ${\rm\bf P}$  and its inverse indeterminate. 
One can thus no longer diagonalize the system of equations as in~\re{diagonal} 
and come up with solutions for~\re{taylor} given by~\re{pert_soln}.

We devise the following scheme to assess stability: 
we introduce small perturbations $\delta{\mathbf{V}}$
around an arbitrary point $\mathbf{V}_{0}$;
using~\re{taylor} and diagonalizing ${\rm\bf M}$, 
we solve for $\delta{\mathbf{V}}$. 
In order for the fixed point to be an attractor we require the following conditions to be satisfied. \textit{If $\mathbf{V}_{0}$ is infinitesimally close to the fixed point $\mathbf{V}_{FP}$,
			then, as $N\to\infty$}:

\begin{itemize}
    \item[(A)]  	 \textit{the perturbations $\delta{\mathbf{V}}$ are infinitesimally close to zero, 
			}
			therefore we require
			\begin{eqnarray}
			\lim_{\mathclap{\substack{N\to+\infty \\ V_{0}\to V_{FP}}}}\quad \mathbf{\delta V} \to 0 \label{deltaV}\, ;
			\end{eqnarray}

    \item[(B)]  	\textit{the derivatives of perturbations, $d\, \delta{\mathbf{V}}/dN$, are infinitesimally close to zero,}
            therefore we must verify that
			\begin{eqnarray}
			\lim_{\mathclap{\substack{N\to+\infty \\ V_{0}\to V_{FP}}}}\quad \frac{d}{dN} \mathbf{\delta V} \to 0 \label{dNdeltaV}\, ;
			\end{eqnarray}
			
    \item[(C)]  \textit{$y/G_{2}(X)$ is infinitesimally close to its fixed-point limit \re{yg2} --}
			in compact form this  translates to
			\begin{eqnarray}
			 \label{ybyg2}
			\lim_{\mathclap{\substack{N\to+\infty \\ V_{0}\to V_{FP}}}}\quad \left(\frac{y}{G_{2}(X)}\right)^{2} \to \qquad\qquad\qquad\qquad\qquad \, \\
			1-\alpha_{\sigma}\frac{G_{1}(X_{FP})(1-\nicefrac{\omega}{6})(G'_{1}(X_{FP}))^{2}}{\left[X_{FP}G'_{1}(X_{FP})(1-\nicefrac{\omega}{6})+\omega G_{1}(X_{FP})\right]^2}\, .  \nonumber
			\end{eqnarray}
\end{itemize}

In more detail, after diagonalizing the matrix \textbf{M}, from~\re{taylor} we obtain
\begin{eqnarray}
    \frac{d}{dN} \mathbf{\delta \bar{V}}=\mathcal{D}\mathbf{\delta \bar{V}}+ \mathbf{B},
    \label{taylor2}
\end{eqnarray}
where
\begin{eqnarray}
    &\mathbf{B}=P^{-1}\mathbf{f}\left(\mathbf{V}_{0} \right),\label{B}\\ \nonumber
\end{eqnarray}

Solving~\re{taylor2}, we are left with
\begin{align}
    \mathbf{\delta \bar{V}}&=\text{Diag}\left(-\frac{1}{\lambda}_{1},\cdots,-\frac{1}{\lambda}_{5}\right)\mathbf{B}\\ \nonumber
    &+\text{Diag}\left(e^{\lambda_{1} N},\cdots,e^{\lambda_{5} N}\right){\mathbf{C}}\,,
\end{align}
where $\mathbf{C}$ is the integration constant vector.

Upon returning to the original basis, we find
\begin{align}
    \label{perturbed}
    \mathbf{\delta {V}}&={\rm\bf P}\hspace{1mm}  \text{Diag}\left(-\frac{1}{\lambda}_{1},\cdots,-\frac{1}{\lambda}_{5}\right){\rm\bf P}^{-1}\mathbf{f}\left(\mathbf{V}_{0}\right)\\ \nonumber
    &+{\rm\bf P}\hspace{1mm}\text{Diag}\left(e^{\lambda_{1} N},\cdots,e^{\lambda_{5} N}\right){\mathbf{C}}.
\end{align}

In Appendix~\ref{ModifiedStability} we show that the requirement (A) could be satisfied if the eigenvalues $\lambda_{i}$'s are either real and negative 
or complex with negative real part. It is also shown that, if $X_{0}$ is a point infinitesimally close to its fixed-point value $X_{FP}$, it must satisfy the following relations\footnote{
In special cases where $G_1(X_{FP})=0$, such as $\omega=6$, or $2\alpha_3^2=\alpha_4$, the eigenvalues involve ratios of zeros that we are unable to resolve, 
so we cannot determine the stability.}
\begin{enumerate}
    \item $\alpha_{\sigma}>0$
    \begin{eqnarray}
G'_{1}(X_{0})G_{2}(X_{0})G'_{2}(X_{0})&>&0\label{reln1},\hskip 2pt 0<\omega<6 \\
G'_{1}(X_{0})G_{2}(X_{0})G'_{2}(X_{0})&<&0\label{reln2},\hskip 2pt \omega>6 \, ,
\end{eqnarray}
\item $\alpha_{\sigma}<0$
    \begin{eqnarray}
G'_{1}(X_{0})G_{2}(X_{0})G'_{2}(X_{0})&<&0\label{reln3},\hskip 2pt 0<\omega<6 \\
G'_{1}(X_{0})G_{2}(X_{0})G'_{2}(X_{0})&>&0\label{reln4},\hskip 2pt \omega>6\, .
\end{eqnarray}
\end{enumerate}
In the numerical investigation performed in Section \ref{scanning}, we will take $X_{0}$ as the initial condition for the dynamical variable $X$. Therefore, relations \re{reln1}-\re{reln4} provide the viable initial conditions needed to have linear stable solutions for the dynamical variables. We note here that \re{reln1}-\re{reln4} can only be satisfied for $X_{0}$ either greater or less than $X_{FP}$ and never both\footnote{We exclude particular cases for which in a neighborhood of $X_{FP}$ not only $G_{2}(X)$ changes sign but also either $G'_{2}(X)$ or $G'_{1}(X)$.}.

For $\mathbf{V_{FP}}$ to be an attractor, we are required to verify under which conditions (B) and (C) are satisfied. However, as explained in more details in Appendix~\ref{ModifiedStability}, in linear perturbation theory, (B) and (C) cannot be determined. 
Therefore a numerical investigation on both \re{dNdeltaV} and \re{ybyg2} should be performed. In practice the analysis of $\left(\frac{y}{G_{2}(X)}\right)^{2}$ is enough. Indeed we find that, even though the dynamical variables approach their fixed-point values in the asymptotic future, $\left(\frac{y}{G_{2}(X)}\right)^{2}$ oscillates with maxima that \textit{grow} in time. Eventually, it reaches the critical value of 1, making $r$ in eq.~\re{rdef} singular. 
We discuss this phenomena in greater detail in Section~\ref{scanning}.
 
 \subsection{Case 2: $\left[\frac{\dot{\phi}^0}{n}\right]_{FP} =0$}

In this Case, all elements of the stability matrix 
are well defined and finite in the fixed-point limit. 
Hence we use the standard approach of fixed-point analysis. 
The analytical form of the elements of ${\rm \bf M}$ around the fixed-point solutions 
are provided in Appendix~\ref{StandardStability} (\re{M22}-\re{M33}). 
For the fixed points to be attractors, 
the elements of ${\rm\bf M}$ must satisfy the conditions~\re{stable} for stable solutions and~\re{stablespiral}
for stable spiral solutions.

\subsection{Case 3: $X_{FP}=0$}
Recalling~\re{sigdotx0}, we find that the presence of terms $\frac{X}{\tilde{\Omega}_{\sigma}}$ and $\frac{y}{G_{2}(X)}$ in the stability matrix make the elements~\re{MM22},~\re{MM23},~\re{MM31},~\re{MM32} and~\re{MM33} and consequently the eigenvalues~\re{eigenvalues} indeterminate. 
Hence the stability of $X_{FP}=0$ is unclear.

\section{ Numerical Investigation of fixed-point stability}
\label{scanning}

As noted above, 
fixed-point linear stability conditions are necessary but not sufficient 
to guarantee that a fixed point can be reached by evolving from an initial configuration
that is displaced from that fixed point. Moreover, as described above, for Case 1 and Case 3 some or all the relevant quantities that appear in the background equations cannot be analytically assessed in linear perturbation theory.
Therefore we need to perform numerical tests of the fixed-points stability.

For suitable values of the parameters 
$(\alpha_{3},\alpha_{4},\omega,\alpha_{\sigma})$,
we check numerically that the 5 dynamical variables approach their fixed-point values
if initially perturbed from them. 
Crucially, this includes verifying that $z=y/G_{2}(X)$ approaches
the value given by~\re{yg2}.

In our numerical investigations,  we use the results of Section~\re{stability} to guarantee  linear stability. We look further for values of the parameters for which:
\begin{enumerate}[(i)]
    \item $X_{FP}\geq0$; \label{check1}
    \item $\tilde{h}_{FP}^2$ given by~\re{hsqfp} is positive;
    \label{check2}
    \item the fixed-point linear stability conditions 
    (given by~\re{reln1}-\re{reln4} for Case 1, and~\re{stable} or~\re{stablespiral} for Case 2) are satisfied.
    \label{check3}
\end{enumerate}

Note that we could further constrain the parameter space by selecting regions where the scalar, vector and tensor perturbations are stable~\cite{AdF-SM,Heisenberg:2015voa,Gumrukcuoglu:2016hic}. 
Although it is not necessary for the present analysis,  to simplify our search, we use some of the necessary restrictions imposed by the stability of the perturbations, which we summarize in Appendix~\ref{perturbations}.   

As discussed below, we find that the traditional (Case 1) fixed points cannot be reached from any neighboring configuration because $z$ does not approach the corresponding fixed-point value!  The Case 2 fixed points behaved as expected from the linear stability analysis. The Case 3 fixed points always encounter a singularity before reaching the fixed-point values.

\subsection{Case 1: $J(X_{FP})=0$}
\label{numCase1}
Our goal in this section is nominally 
to identify values of the parameters $({\alpha_{3},\alpha_{4},\omega,\alpha_{\sigma}})$ 
such that~\re{check1}-\re{check3}  hold true 
and verify that the dynamical variables approach 
their fixed-point values in the asymptotic future. 
One of the central results of this paper is that we fail to do so. 
We show that, for all the choices of parameters and initial conditions we consider, 
the dynamical variables {\em never} approach their fixed-point values
if they do not start there to begin with.

We begin the discussion with an example by setting 
$({\alpha_{3},\alpha_{4},\omega,\alpha_{\sigma}})=(3.1, 3.1, 5.5, 10.0)$. 
Solving~\re{xpm},
we find that $X_{+}$ satisfies the conditions~\re{check1}-\re{check2}. This set of  parameter values also satisfies   the 
tensor, vector and scalar  stability  at the $X_{+}$ fixed point   (see Appendix ~\ref{perturbations}).  

We remind the reader that since $G_{2}(X)$ vanishes at the fixed point, the stability conditions~\re{reln1}-\re{reln4} only hold true in the vicinity of the fixed point. In this example, since $0<\omega<6$ and $\alpha_{\sigma}>0$, the initial value of $X$ must be of the form $X_{0}=X_{+}+\epsilon$, where $0<\epsilon\ll1$ so that~\re{reln1} is satisfied. To study the behavior of the dynamical variables close to the fixed point, we set $\epsilon={10^{-6}}$ and the initial conditions to be ${\tilde{\Omega}}_{DE, 0} = {\tilde{\Omega}}_{DE, FP}+\epsilon$, $ {\tilde{\Omega}}_{m, 0}=\epsilon$, ${\tilde{\Omega}}_{r, 0}=\epsilon$.
We recall that for Case 1, we cannot freely perturb $y$, 
because the initial value of $r$ from~\re{rdef} would  not then 
be close to its fixed-point value~\re{rfp}. 
Instead, we perturb $r$ by $\epsilon$ and initialize $y$ using~\re{yini}. 
For a detailed discussion, we refer the reader to Appendix~\ref{ModifiedStability}. 
Notice that ${\tilde{\Omega}}_{DE, 0}$, $r_{0}$, $y_{0}$ could be either greater or less than the fixed-point values. In this example we chose the former.

\begin{figure}
\centering
\includegraphics[width=1\hsize]{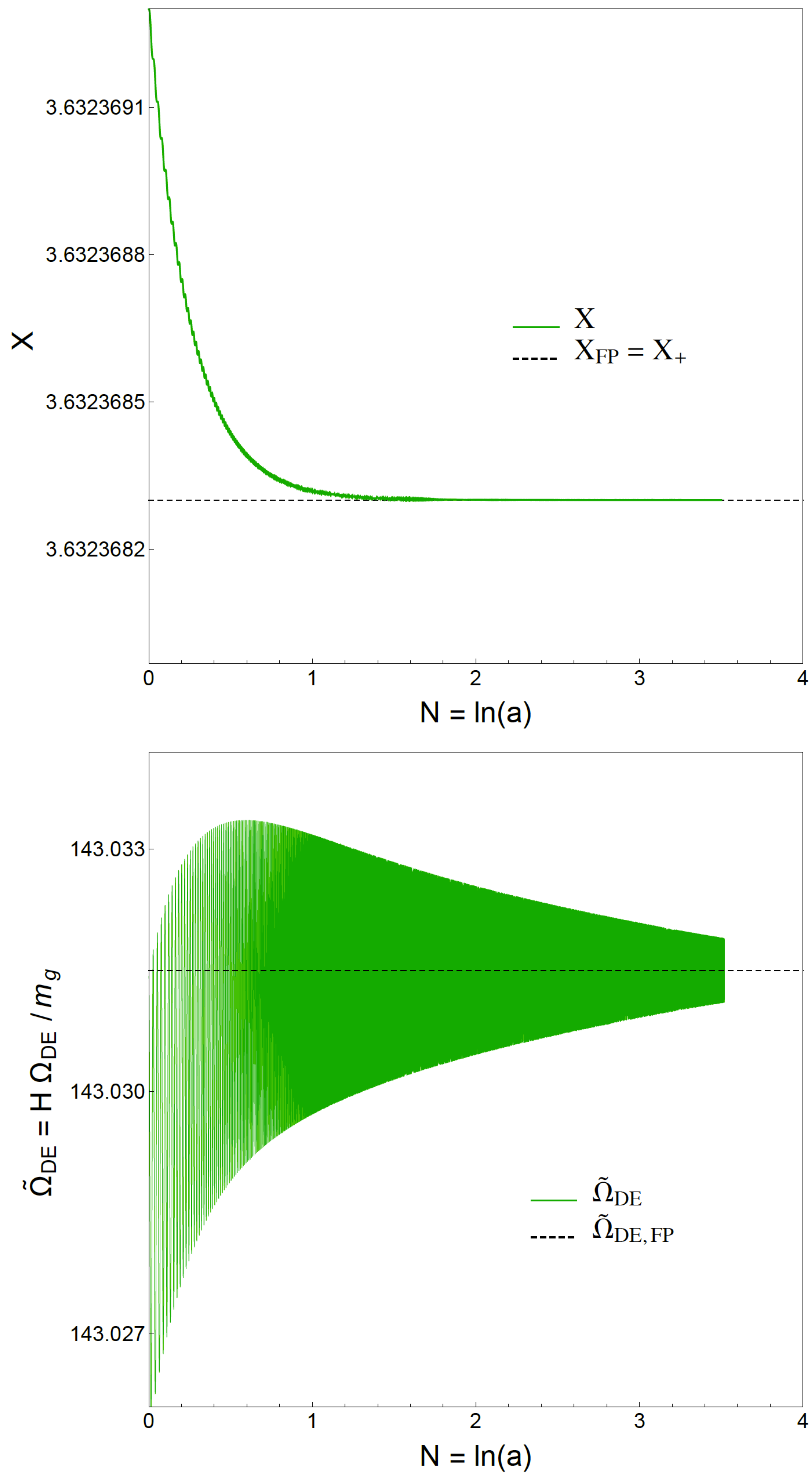}
\caption{
\label{dyn_var} $X(N)$ ({\it top panel}) and ${\tilde{\Omega}_{DE}}(N)$ ({\it bottom panel}) starting from initial values of dynamical variables $\epsilon = 10^{-6}$ away from their fixed-point values. The parameters are set to $(\alpha_{3},\alpha_{4},\omega,\alpha_{\sigma})=(3.1,3.1,5.5,10.0)$. 
$X$ and ${\tilde{\Omega}_{DE}}$ approach their fixed-point values until they hit singularities.}
\end{figure}

After setting the initial conditions as described above, we study the behavior of the dynamical variables with time.
Equations~\re{yN},~\re{matter} and~\re{radiation} have simple solutions, 
\begin{eqnarray}
\label{yOmOr}
y=y_{0}e^{-4N},\hskip 2pt\tilde{\Omega}_{m}=\tilde{\Omega}_{m,0}e^{-3N},\hskip 2pt\tilde{\Omega}_{r}=\tilde{\Omega}_{r,0}e^{-4N}\,.
\end{eqnarray} 
The evolution equations for $X$ and $\tilde{\Omega}_{DE}$ have no analytical solutions. 
Therefore, we use~\re{yOmOr} in~\re{hsqnew} and~\re{rdef}, 
and solve equations~\re{txx} and~\re{tildeOmsigmaeqs} numerically. 
We find that~\re{txx} and~\re{tildeOmsigmaeqs} evolve until they reach a singularity. 
As shown in Fig.~\ref{dyn_var}, 
the evolution of $X$ and $\tilde{\Omega}_{DE}$ before the singularity 
turns out to be exactly what we expect from perturbation theory: 
they oscillate around their fixed-point values with decaying amplitude and increasing frequency.

\begin{figure}
\centering
\includegraphics[width=1\hsize]{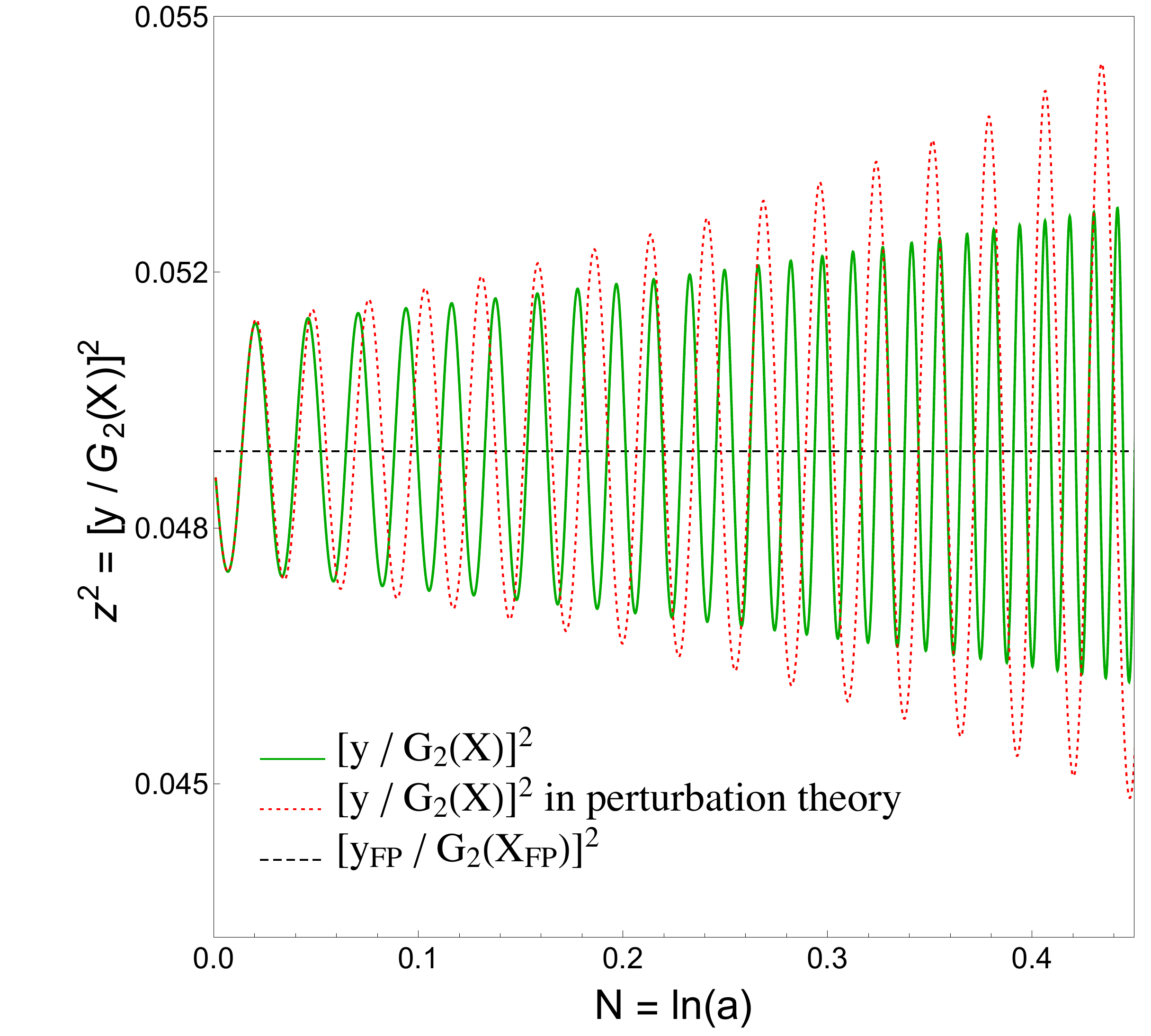}
\caption{\label{zsq_both}
$z^2(N)$ evaluated from solving the full equations (non-perturbative) and and linear perturbation theory. Parameter values and initial conditions are the same as for Fig.~\ref{dyn_var}.}
\end{figure}

\begin{figure}
\centering
\includegraphics[width=1\hsize]{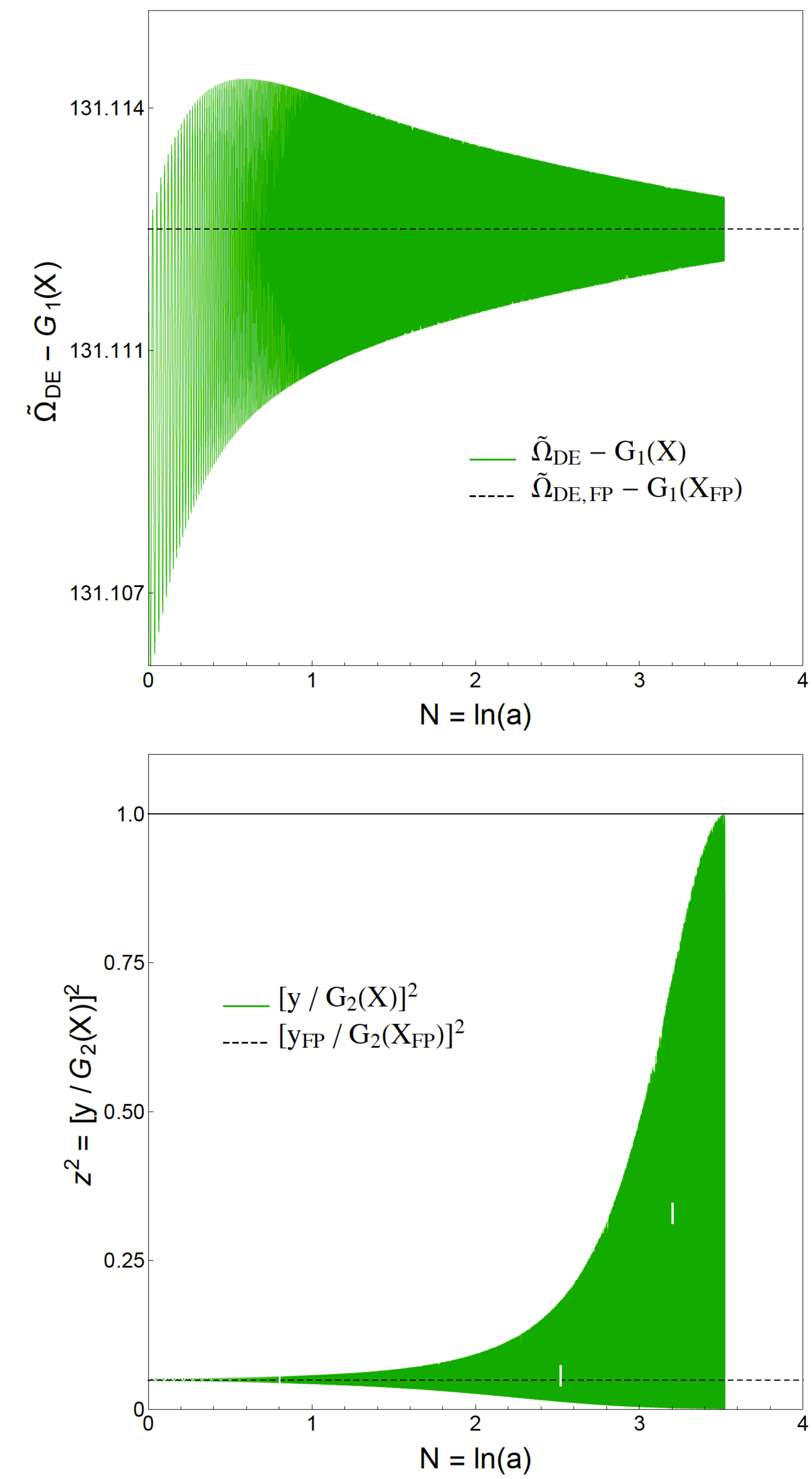}
\caption{\label{zsq_full}
$\left(\tilde{\Omega}_{DE}-G_{1}(X)\right)$ ({\it top panel}) and $z^{2}$ ({\it bottom panel}) in function of $N$. $\left(\tilde{\Omega}_{DE}-G_{1}(X)\right)$ is well behaved. $z^{2}$ shows oscillations with peaks that grow in time and eventually reach unity making the equations singular. 
}
\end{figure}

By inspecting the RHSs of ~\re{txx} and~\re{tildeOmsigmaeqs} we can identify the possible sources of the singular behavior. 
The only possibilities are that, as the fixed point is approached, the square-root term
$\left(\tilde{\Omega}_{DE}-G_{1}(X)\right)$ in~\re{txx} or $\left(1-\left(\frac{y}{G_{2}(X)}\right)^{2}\right)$
 in ~\re{tildeOmsigmaeqs} approaches zero. From the {\it top panel} of Fig.~\ref{zsq_full} we see that $\left(\tilde{\Omega}_{DE}-G_{1}(X)\right)$ does not vanish as the fixed point is approached. 
 On the other hand in Fig.~\ref{zsq_both} we show that $z^{2}=\left(\frac{y}{G_{2}(X)}\right)^{2}$ exhibits an oscillating behavior where the peaks grow in time monotonically.
Thus $z$ does not approach its expected fixed-point value. Moreover, as shown in the {\it bottom panel} of Fig.~\ref{zsq_full}, it eventually approaches unity, 
causing a breakdown in the coupled dynamical
equations for $X$ and $\tilde{\Omega}_{DE}$. 

To ensure that the phenomenon described above is not a numerical artifact, we performed
numerous numerical and analytic tests.
Numerically, we confirmed that the $z^2\to1$ behavior was robust to: 
increasing  the numerical precision demanded from the integrator, 
increasing the order of the integrator,
changing the integration scheme,
and 
changing the initial conditions. Therefore, we have conclusive evidence that the evolution of the dynamical variables drives $z^{2}$ towards unity. 
We therefore conclude that for this value of the EQDMG parameters,
in the neighborhood of a Case 1 fixed point
the evolution equations~\re{txx} and~\re{tildeOmsigmaeqs} 
reach a singularity in the asymptotic future. 

In order to confirm that this conclusion holds true for all regions of the parameter space we would need to evolve the differential equations for a high number of different values of $({\alpha_{3},\alpha_{4},\omega,\alpha_{\sigma}})$ and initial conditions close to the fixed-point solutions.
This is numerically too costly, therefore we solve the exact equations for a limited number of points in the parameter space (few thousands). For a more extensive scan (millions of points) we rely on a sensible approximation based on perturbation theory and on the following argument. 

If the fixed point is an attractor, 
perturbation theory should be increasingly accurate 
the closer the initial conditions are to the fixed-point value;
if the fixed point is not an attractor, then perturbation theory may or may not work.
If in perturbation theory we were to find that $z$  approached its fixed-point value, 
then we would want to verify that conclusion with the full non-perturbative solution. 
If $z$ does not approach its fixed-point value in perturbation theory,
then that may be because the initial perturbation away from the fixed point was too large 
-- outside the basin of attraction --
so one should decrease the magnitude of the perturbation as much as possible.
The smaller the perturbation, 
the more one would expect to trust  perturbation theory.
If this still fails,
it is always possible that the basin of attraction is extremely small
and difficult to find numerically; but in any case, while
we will not have arrived at a mathematical proof that this fixed point is not an attractor,
we will certainly have shown that it is not  a suitable candidate for a cosmological model.

We start by testing the prediction of $z^{2}$ from perturbation theory in the previously considered example. We rely on equations \re{dely}, \re{delX}, \re{delDE}, \re{delm}, \re{delr} where the $C_{i}$ coefficients are determined by requiring that $\mathbf{\delta {V}}(N_{in}\equiv0)=0$. In Fig.~\ref{zsq_both} we plot the behavior of $z^{2}$ for the first $\sim{20}$ periods comparing the full and perturbation theory solutions. We verify a consistent growth in the peaks of $z^{2}$. Hence, in the subsequent analysis, we employ the following supporting argument: if in perturbation theory the first $20$ peaks of $z^{2}$ grow monotonically with time, we conclude that $z^{2}$ does not approach its fixed-point value. We randomly selected $\sim{10^7}$ values of the parameters in the following ranges 
\begin{align}
\label{region2}
10^{-5}&<\left|\alpha_{3}\right|,\left|\alpha_{4}\right|,\alpha_{\sigma}<10^{5},\nonumber \\
0&<\omega<6 \text{  or  } 6<\omega<10^{5}
\end{align}
and evaluate $X_{\pm}$. We evaluate the first $\sim{20}$ peaks of $z^2$ for all points that satisfied the conditions~\re{check1}-\re{check3} together with the  pertinent ones in Appendix~\ref{perturbations} using linear perturbation theory. We set the initial conditions of the dynamical variables to be from $\epsilon={10^{-8}}$ to $\epsilon={10^{-12}}$ away from their fixed-point values\footnote{For simplicity we only consider ${\tilde{\Omega}}_{DE, 0}$, $r_{0}$, $y_{0}$ larger than their fixed-point values. We numerically inspect few examples with initial conditions smaller than the fixed-point values and we find the same singular behavior.}. 
We repeated the same procedure 
using the numerical solutions to the full (non-perturbative) equations
with  few thousand randomly selected points.
We find that the peaks of $z^2$ always grow monotonically 
in both linear perturbation theory and using the exact equations. 

Notice that for $\alpha_{\sigma}<0$, equations~\re{defn} and \re{ydef} imply $z^{2}>1$. However, even for negative values of $\alpha_{\sigma}$, $r$ from~\re{rdef} still becomes singular because the troughs of $z^{2}$ monotonically decrease and approach unity.

Hence we come to the following conclusion: 
\textit{the standard (Case 1) fixed-point solutions 
of  Extended Quasi-Dilaton Massive Gravity 
are dynamically unattainable 
due to an unavoidable singularity while approaching them. 
Hence, we rule out the suitability of this fixed point
as a cosmological model.} 

\subsection{Case 2: $\left[\frac{\dot{\phi}^0}{n}\right]_{FP} =0$}
Recalling that $\alpha_{\sigma}$ is positive,  
we randomly select points in the ranges~\re{region2}. 

To obtain the fixed-point values of $X$, we use~\re{7th} and~\re{hsqfp} to arrive at a 7th order polynomial in $X$ with no analytical solutions. After numerically evaluating the roots of the polynomial, we demand that the fixed-point values satisfy~\re{check1}-\re{check3} and that the vector and tensor perturbations are stable\footnote{We keep this analysis agnostic to the stability conditions for scalar perturbations since, as we show below, the fact that $\dot{\phi^0}=0$ does not guarantee the absence of the BD ghost.}(see Appendix~\ref{perturbations} for details). We find parameter values for which both stable and spiral solutions are allowed. We selected a few points in the allowed parameter space to verify the attractor behavior of the fixed points. Starting from small perturbations ($\epsilon =10^{-6}$) around the fixed-point values of the dynamical variables, we study the evolution of the differential equations with time by solving~\re{yN},~\re{matter} and~\re{radiation} analytically and~\re{txx} and~\re{tildeOmsigmaeqs} numerically. As predicted from linear stability analysis (Section \ref{stability}), we find that the dynamical variables reach their fixed-point values in the asymptotic future.

\subsection{Case 3: $X_{FP}=0$}

For the Case 3 fixed points we do not have indications from the linear stability analysis. We perform a random search selecting $64\times10^{4}$ points in the parameter space. Starting from small perturbations ($\epsilon =10^{-6}$) around the fixed-point values we find that the dynamical variables are approaching the fixed-point limit. However, we find that the quantity $\left(\tilde{\Omega}_{DE}-G_{1}(X)\right)$ always approaches zero at finite time. We find numerically that this makes $r$ vanish which is not allowed in the theory.

\section{Perturbations: Avoiding the Boulware-Deser Ghost}
\label{bdghost}

One of the outstanding concerns in any theory that adds a scalar degree of freedom
to Einstein-Hilbert gravity is the possibility that the theory includes 
a Boulware-Deser ghost --
 a dynamical  degree of freedom with a wrong-sign kinetic term.

Due to disagreements in the literature  mentioned above (see \cite{Kluson:2013jea,Mukohyama:2013raa,Heisenberg:2015voa}), 
in this section we reconsider the analysis of the scalar perturbations,
and determine under what conditions there is or not a necessary additional constraint equation that generically allows 
one to eliminate the Boulware-Deser mode.

We will show that such a constraint exists when $J(X)=0$ (Case 1) or $\dot{\sigma}=0$, giving the possibility of being ghost-free
at the corresponding fixed points.   In both cases, there is also a well-defined set of initial conditions of
the dynamical variables for which this virtue
extends beyond the fixed point. The first case is the only one potentially relevant for cosmology, and we explore the consequences of the  initial 
conditions below.  The second case $\dot{\sigma}=0$ corresponds to a  Minkowski  background metric (provided $X\neq 0$) which we consider only in Appendix \ref{Minkowski}. 

Given that in the previous section, for Case 3 we have shown $X=0$ cannot be approached dynamically from any neighborhood, one might consider the possibility when $X(N)=0$ $\forall$ $\rm{N}>0$. This possibility can also be ruled out because $X=0$ only makes sense as a fixed-point limit. Hence from now on, we will not consider Case 3 any further.

Following the standard treatment and in this section only, we take the action of the  matter sector to be \footnote{Clearly, the action (80) does not describe dust-like matter. However, we consider it only for simplicity because it will be enough for the  purposes of this section.}
\begin{equation}\label{sm}
S_m= \int d^4x \sqrt{-g} \,P(Y,\chi)\,,
\end{equation}  
which corresponds to the addition of a scalar field $\chi$ with a non-canonical kinetic term, given by the function $P(Y,\chi)$, where
\begin{equation}
Y \equiv -\frac{1}{2}\partial_\mu\chi\partial^\mu\chi\,.
\end{equation}
The  fluid  variables (pressure $p$, energy density $\rho$ and sound speed $c_s$) 
associated with $\chi$ can be written as \cite{ArmendarizPicon:2000ah}:
\begin{equation}
 p = P (Y,\chi) \,,\qquad 
\rho \equiv 2P_{,Y} Y - P, \qquad
 c_s^2 \equiv \frac{P_{,Y}}{\rho_{,Y}}\,.
\label{eq:fluid}
\end{equation}  
To study the perturbations in $\chi$, we replace  $\chi$ by $\chi+M_{\rm Pl}\delta\chi$.

To facilitate comparisons,
we adopt notations as close as possible to those most used in the literature. 
We  decompose the metric into tensor, vector, and  scalar as 
\begin{eqnarray}
\delta g_{00} &=& -2\, \Phi\,,\\
\delta g_{0i} &=&  \,a\,\left(\partial_i B+B_i\right)\,,\nonumber\\
\delta g_{ij} &=& a^2 \left[2\,\delta_{ij}\psi +\left(\partial_i\partial_j-\frac{\delta_{ij}}{3}\partial^k\partial_k\right)E+\partial_{(i}E_{j)}+h_{ij}\right]\,,\nonumber
\end{eqnarray}
where  $\delta^{ij}h_{ij} = \partial^ih_{ij} = \partial^i E_i = \partial^i B_i=0$ and  Latin indices are raised with $\delta^{ij}$. 
For the quasi-dilaton field we write the background plus perturbation
by replacing $\sigma$ by $\sigma +  M_{\rm Pl}\delta\sigma$.

For the sake of clarity, in this section we present only 
those results that are either more generic or different from those in previous literature, 
and relegate the remaining details to Appendix \ref{AppendPerturb}.

We focus first in the case $\dot{\phi^0}\neq 0$  and, following literature, we adopt the unitary gauge $\phi^{a}=\phi^{0}(t)\delta^a_0 +\delta^a_ix^i$.  The case   $\dot{\phi^0}(t)= 0$,  for which we cannot work in this gauge, is analyzed later.

\subsection{$\dot{\phi^0}\neq 0$}
 
The part of the action~(\ref{eq:action})  that is quadratic in the perturbations  
can be split into tensor, vector and scalar contributions. 
We focus here  on the scalar part.

From the variation of the quadratic action for the scalar sector 
with respect to  $\Phi$ and $B$,  
we obtain constraint equations that allow us to eliminate these two variables.  
The solution for  $\Phi$ and $B$ can be found for a generic background 
and without making  use of the background equations\footnote{
    There are some disagreements in the literature regarding these results.  
    In the appropriate limit, our results reduce to those of \cite{Gumrukcuoglu:2016hic} 
    rather than those of \cite{Heisenberg:2015voa}. For more details see Appendix \ref{AppendPerturb}.}.
Introducing the solutions for $\Phi$ and $B$ back into the action, 
we can write the kinetic part as 
\begin{equation}\label{kineticAction}
S^{(2)}_{\rm scalar}\supset 
\frac{M_{\rm Pl}^2}{2}\int d^3k \,dt\,a^3 \dot{Z}^\dagger\,\mathcal{Q}\,\dot{Z}\,,
\end{equation} with $Z=\{\psi,\delta\sigma, E,\delta\chi\}.$ One combination corresponds to the Boulware-Deser mode. For this mode to be non-dynamical, 
the determinant of the matrix $\mathcal{Q}$ must vanish.

We first consider the vacuum case, and then see how the inclusion of matter
changes the results.

\begin{itemize}
 \item {Vacuum case}: In the absence of the additional field $\chi$,
$\mathcal{Q}$ is a $3\times3$ matrix and $Z=\{\psi,\delta\sigma, E\}.$   
The determinant ${\rm Det}(\mathcal{Q})$ can be computed analytically in Fourier space.  
After expanding in powers of  comoving wavenumber $k$,  
keeping only the leading order terms (the infrared part),  
and  using the background equations to express ${H}$ and $\dot{H}$ 
in terms of the other dynamical variables, 
\begin{align} 
   & {\rm Det}(\mathcal{Q})=\\ \nonumber
    &\frac{4 \omega \alpha_\sigma M_{\text{Pl}}^4 a^2 J(X) {\dot{\phi^0}}^2 \dot{\sigma}^2  k^4}
     {r^3 X m_g^2 \left(J(X)+\alpha_3 (X-1)^2-3 X+3\right)}
     +\mathcal{O}(k^6)\,.\nonumber
\end{align} 
The fact that the background equations are satisfied 
does not guarantee that the determinant vanishes,
as previous computations suggested \cite{Heisenberg:2015voa}. 
However, it is clear that ${\rm Det}(\mathcal{Q})$ vanishes when  
$J(X)=0$ or ${\dot{\sigma}}=0$.   Moreover, in that cases it can be shown that the determinant vanishes at all order in $k$.
As we show next, these results are robust to the addition of matter.

\item {With Matter}:
Proceeding as above, 
at leading order in the wavenumber $k$, 
after  using  the background equations to replace 
$\dot{H}$ and   ${H}$ in terms of the other dynamical variables, 
the determinant of the now $4\times 4$  matrix $\mathcal{Q}$  can be written as  
\begin{eqnarray} 
 &&{\rm Det}(\mathcal{Q})=  
    \frac{32 \omega }{     r^3 X } \alpha_{\sigma } M_{\text{Pl}}^6 a^2 J(X)\dot{\phi^0}^2 (X-1) P_{,Y}(Y,\chi) 
    {\dot{\sigma}}^2 k^4  \nonumber\\
     &&\times       \Big{\{}p+\rho(1-2 c_s^2)-2 c_s^2 (X-1) m_g^2 M_{\text{Pl}}^2 \nonumber\\
      &&\times \left(J(X)+\alpha _3 (X-1)^2-3 X+3\right)   \Big{\}}^{-1}     +\mathcal{O}(k^6)\,.
\end{eqnarray}    
Therefore,  ignoring the particular cases for which $X=1$, 
we see that the determinant vanishes in the infrared under the same conditions as in the vacuum case. Under those conditions, we also check the determinant vanishes at all order in $k$.

Of course the determinant also vanishes in the case where the matter is just a
 cosmological constant, $P_{,Y}=0$.  
 
\end{itemize}

Therefore,   ${\rm Det}(\mathcal{Q})=0$ at the Case 1  fixed points.
If we perturb $X$ away from that fixed-point value, then $J(X)\neq0$,
and ${\rm Det}(\mathcal{Q})\neq0$ either. 
 However, from~\re{Constraint},
we can  conclude that if $J\equiv J(X)=0$ at some initial time $t_0$,
then also $\dot{J}=0$ at that time, and also $\ddot{J}$.  
Thus $J(X)$ remains zero at all times once $X=X_{FP}$.  
Consequently ${\rm Det}(\mathcal{Q})$ also remains equal to zero. 
Therefore, in order to eliminate the BD ghost in both the vacuum and matter contexts,  we must impose special initial conditions on the dynamical variables; namely, that  $X=X_\pm$ is  exactly satisfied.
Setting $X$ to this fixed-point value, and thus setting $J(X)=0$, $y=0$ and $\dot{X}=0$,
we can solve~\re{matter} through~\re{rdef} to obtain:
\begin{eqnarray}
\label{jxcond}
\tilde{\Omega}_{DE}\ &=& \frac{\frac{\omega}{6}\left(\tilde{\Omega}_{m0}e^{-3N}+\tilde{\Omega}_{r0}e^{-4N}\right)+ G_{1}(X_\pm)}{1-\frac{\omega}{6}}\nonumber\\
\tilde{h}^2 &=& \frac{6}{\omega}\left(\tilde{\Omega}_{DE}-G_1(X_\pm)\right)
\end{eqnarray}
This appears to be just a fine-tuning of the dynamical variable $X$ 
to some parameter-dependent value; however $X$ is a function of 
both the quasi-dilaton and the scale factor given by~\re{xsigma}.
This thus appears to be an awkward fine-tuning, relating the initial
values of many of the dynamical variables to one another. It also allows the fixed points to become stable attractors that can be approached dynamically. Furthermore, notice that in the asymptotic past, $N\to-\infty$, the matter and radiation terms will dominate over $G_{1}(X\pm)$. This restricts the values of $\omega$ to $0<\omega<6$ or it means that the theory is not valid arbitrarily far into the past.

\subsection{$\dot{\phi^0}= 0$}

As mentioned earlier, in the special case $\dot{\phi^0}= 0$ (Case 2) we cannot use the unitary gauge.
 Assuming $H\neq 0$ we choose the gauge with $\psi=0$ instead, while we keep $\phi^{i}=x^i$.    
In principle,  as first noticed by~\cite{Gumrukcuoglu:2016hic},  this case could also be interesting beyond  the fixed-point limit. 
This corresponds to setting  $\dot{\phi^0}(t_0)= 0$ at some initial time $t_0$, yielding  $\ddot{\phi^0}(t_0)= 0$ and  $\dot{\phi^0}(t)= 0$ for all time. This enforces $y=0$, but the other dynamical variables remain free to evolve.

The perturbations of the St\"uckelberg  scalar degree of freedom $\delta\phi^0$ (since $\dot{\phi^0}=0$ at the background level) only enter as a contribution to $f_{\mu\nu}$ that is quadratic in $\delta\phi^0$. Therefore, the quadratic action for this  scalar decouples from the other parts, and the kinetic part can be immediately computed  
\begin{equation}
S^{(2)}_{\rm scalar} \supset - \frac{M_{Pl}^2}{2} \int d^3k \,dt\,a^3 \,\frac{m_g^2}{r} G_2(X) a^2|\delta\dot{\phi^0}|^2\,.
\end{equation} 

Now, to compute the determinant of the kinetic matrix corresponding to the other degrees of freedom we proceed  as above.  We  integrate out $\Phi$ and $B$, we  write the relevant part of the action as in (\ref{kineticAction}), and   we consider the vacuum case and the case with matter  separately:
 
\begin{itemize} 

\item Vacuum case: In this case  $Z=\{\delta\sigma,E\}$, and 
\begin{eqnarray}
&&{\rm{Det}}(\mathcal{Q})=G_1'(X) k^4 m_g^2 X \omega  a^2 \left(6 \alpha_{\sigma} G_1(X)+r^2 X^2 \omega
   \right)\nonumber\\
&&\times\Big{\{} 4 \left(9  \alpha_{\sigma}G_1(X) G_1'(X) m_g^2 X a^2+k^2 (r+1) \left(6
    \alpha_{\sigma} G_1(X)\right.\right.\nonumber\\
    &&    +r^2 X^2 \omega ))\Big{\}}^{-1};
    \end{eqnarray}
    
   \item With matter: $Z=\{\delta\sigma,E,\delta\chi\}$, and
\begin{eqnarray}
&&{\rm{Det}}(\mathcal{Q})=-H^2 k^4 X \omega  m_g^2 G_1'(X) M_{\text{Pl}}^2\nonumber \\
&&\times\left[X m_g^2 \left(r^2 X \omega -(r-1) \alpha _{\sigma } G_1'(X)\right)+2 \dot{H} \alpha _{\sigma
   }\right]\nonumber\\
   &&\times
 \Big{\{}\dot{\chi }^2 \left[a^2 X m_g^2 G_1'(X) \left(2 \alpha _{\sigma } \left(3 H^2 c_s^2+\dot{H}\right)\right.\right.\nonumber\\
 &&-\left.\left.r^2 X^2 \omega  \left(c_s^2-1\right)
   m_g^2\right)-a^2 (r-1) X^2 \alpha _{\sigma } m_g^4 \left(G_1'(X)\right){}^2\right.\nonumber\\
&   &+\left.4 H^2 k^2 (r+1) \alpha _{\sigma } c_s^2\right]\Big{\}}^{-1}
\end{eqnarray} 

\end{itemize} 
Hence, we   conclude that  the condition $\dot{\phi^0}=0$ is not sufficient to obtain the required additional constraint.

\section{Conclusions}
\label{cnclsns}
In this paper we studied the fixed-point solutions of EQDMG in great detail after splitting them into four cases. 

We performed a linear stability analysis of the background (homogeneous) fixed-point solutions.
This stability analysis for the standard case (Case 1) 
required an unconventional approach. 
We derived necessary stability conditions for the dynamical variables. However, we verify numerically
that the dynamical variables inevitably encounter a singularity 
while approaching their fixed-point values. 
This is because a function of two of the dynamical variables fails
to approach its fixed-point limit, and instead oscillates, 
finally reaching a critical value at which the dynamical equations are no longer valid.

On the other hand, in Case 2,
the dynamical variables smoothly asymptote towards their  fixed-points values. However, in this case, the presence of the additional constraint  that would allow one to eliminate the BD mode is not guaranteed.
A numerical search of the Case 2 parameter space revealed many values of the parameters
that have stable evolutions toward the fixed point.

We analyzed the conditions under which the   constraint  equation that generically allows for the elimination of the  Boulware-Deser ghost can be obtained.
We found that the constraint equation  exists in Case 1 type of solutions, but 
not  around a generic background. Moreover, in these solutions the fixed points are attractors. However,  such solutions require  the time-derivative of the quasi-dilaton field must be exactly tuned against the Hubble
parameter. 

The conclusive result of our study indicates that the EQDMG theory shows pathological behaviors when a generic FLRW solution is assumed. Only an ``awkward" fine-tuned solution is healthy. From our point of view this finding makes the EQDMG less appealing as a viable model to explain the evolution of our Universe.

The extensive analytical and numerical analysis presented for the EQDMG theory must be carried out for all the proposed massive gravity theories that provide flat Friedmann-Lema\^\i tre-Robertson-Walker solutions (a non-exhaustive
list includes \cite{Kluson:2013jea, 2014JCAP...12..011M, 2013PhRvD..88l4006D, 2012PhRvD..86l4014H, 2014IJMPD..2343006D}).

As mentioned below~\re{jxcond}, the condition $\omega<6$ arises in EQDMG when the $J(X)=0$ solution is taken to describe the past cosmological evolution, when matter and radiation dominated.  In EQDMG  this solution is imposed to avoid the presence of the BD ghost. However, since   in   QDMG the BD mode  can be eliminated without  setting $J(X)=0$, the  restriction $\omega<6$ is in principle unnecessary.     It would be worth exploring the   parameter region $\omega>6$  to  see whether the simpler QDMG theory allows a proper description of the expansion history of the Universe.  We propose to contribute to the understanding of this issue in the future.

\acknowledgements
We thank especially C.~de Rham for helpful insights and M.~Fasiello, K.~Hinterbichler, A.~Tolley and S.-Y.~Zhou  for discussions. SA, SK and GDS are partially supported by CWRU grant DOE-SC0009946. The research leading to these results has received funding from the European Research Council under the European Community Seventh Framework Programme (FP7/2007-2013 Grant Agreement no. 279954) ERC-StG "EDECS". 
SA and GDS thank the CERN theory group for their hospitality. We acknowledge the use of the xAct - xPand package for Mathematica \cite{2013CQGra..30p5002P}.

\appendix
\section{Stability Matrix}
\label{StandardStability}

This appendix complements the results presented in Section \ref{stability}. In particular we provide, in linear perturbation theory, the full expressions 
for the stability matrix and its eigenvalues.  
We then present the Case 2 fixed-point limit of the matrix elements and the conditions of stability. 

\subsection{General expression}

The derivative/stability matrix ${\rm \bf M}$ defined in~\re{stabmat} takes the form
\[
{\rm\bf M}=
  \begin{bmatrix}
  \label{M}
    -4 & 0 & 0 & 0 & 0 \\
    0 & M_{22} & M_{23} & M_{24} & M_{25} \\
    M_{31} & M_{32} & M_{33} & 0 & 0 \\
    0 & 0 & 0 & -3 & 0 \\
    0 & 0 & 0 & 0 & -4
  \end{bmatrix}
\]
where
\begin{equation}
\label{MM22}
M_{22}=-1\pm\sqrt{\frac{6\tilde{\Omega}_{\sigma}}{\htildesq\omega}}\left(1-\frac{X G_{1}'(X)}{2\tilde{\Omega}_{\sigma}}\right)\, ,
\end{equation}
\begin{equation}
\label{MM23}
M_{23}=\pm\frac{\sqrt{\frac{3}{2}}X\left(\htildesq-\tilde{\Omega}_{\sigma}\right)}{\sqrt{\tilde{\Omega}_{\sigma}\tilde{h}^6 \omega}}\, ,
\end{equation}
\begin{equation}
\label{MM24}
M_{24}=\mp X\sqrt{\frac{3 \tilde{\Omega}_{\sigma}}{2\tilde{h}^6\omega}}\, ,
\end{equation}
\begin{equation}
\label{MM25}
M_{25}=M_{24}\, ,
\end{equation}
\begin{equation}
\label{MM31}
M_{31}=\frac{\sqrt{\frac{6\alpha_{\sigma}\tilde{\Omega}_{\sigma}}{\omega \left(1-\left(\frac{y}{G_{2}(X)}\right)^{2}\right)}}G_{1}'(X)}{y\left(1-\left(\frac{y}{G_{2}(X)}\right)^{2}\right)}\left(\frac{y}{G_{2}(X)}\right)^{2} \, ,
\end{equation}
\begin{eqnarray}
\label{MM32}
M_{32}&=&-\sqrt{\frac{3}{2}}\sqrt{\frac{\alpha_{\sigma}}{\omega \left(1-\left(\frac{y}{G_{2}(X)}\right)^{2}\right)\tilde{\Omega}_{\sigma}}}(G_{1}'(X))^{2}\, \\ \nonumber
&+&\left(5G_{1}'(X)-M_{31}G_{2}'(X)\left(\frac{y}{G_{2}(X)}\right)\right)\, \\ \nonumber
&+&XG_{1}''(X)\left(-1+\sqrt{\frac{6\alpha_{\sigma}\tilde{\Omega}_{\sigma}}{X^2\omega \left(1-\left(\frac{y}{G_{2}(X)}\right)^{2}\right)}}\right)\, ,
\end{eqnarray}
\begin{equation}
\label{MM33}
M_{33}=-6+\sqrt{\frac{3}{2}}\sqrt{\frac{\alpha_{\sigma}}{\omega \left(1-\left(\frac{y}{G_{2}(X)}\right)^{2}\right)\tilde{\Omega}_{\sigma}}}G_{1}'(X)\, .
\end{equation}
In the above equations, one must replace $\htildesq$ by~\re{hsqnew} and $\tilde{\Omega}_{\sigma}$ by $\tilde{\Omega}_{DE}-G_{1}(X)$.

\subsection{Case 2}
We evaluate ${\rm \bf M}$ for the Case 2 fixed point.

Notice that $\tilde{\Omega}_{\sigma}$, when $X\neq0$ (using~\re{sigdot}), is given by
\be
\tilde{\Omega}_{\sigma}=\frac{\omega \htildesq}{6}\, .
\ee

As explained in Section~\ref{fixedpoint}, in Case 2 we must take the upper sign in \re{txx}, this implies that in \re{MM22}-\re{MM25} the sign is also positive. The elements of the ${\rm \bf M}$ matrix are given by 
\begin{equation}
M_{22}=-\frac{3X G_{1}'(X)}{\htildesq\omega} \label{M22}\, ,
\end{equation}
\begin{equation}
M_{23}=\frac{3X}{\htildesq\omega}\left(1-\nicefrac{\omega}{6}\right)\label{M23}\, ,
\end{equation}
\begin{equation}
M_{24}=-\frac{X}{2\htildesq}\, ,
\end{equation}
\begin{equation}
M_{25}=M_{24}\, ,
\end{equation}
\begin{equation}
M_{31}=0\, ,
\end{equation}
\begin{eqnarray}
M_{32}&=&-3\sqrt{\frac{{\alpha}_{\sigma}}{{\omega}^{2}{\tilde{h}}^{2}}}(G'_{1}(X))^{2}+5G'_{1}(X)\, \\ \nonumber
&+&XG''_{1}(X)\left(-1+\sqrt{\frac{{\alpha}_{\sigma}\htildesq}{X^2}}\right)\, , 
\label{M32}
\end{eqnarray}
\begin{equation}
M_{33}=-6+3\sqrt{\frac{{\alpha}_{\sigma}}{{\omega}^{2}{\tilde{h}}^{2}}}G'_{1}(X)\, ,
\label{M33}
\end{equation}
at the fixed point. 
In the above equations, 
one must\footnote{
    Except in the particular case where $\omega=6$, in
    which case one must instead use a root of  
    \be
    \label{hsqw6}
    \nonumber
    \left(6\htildesq+XG'_{1}(X)\right)^2=\left(G'_{1}(X)\right)^2\alpha_{\sigma}\htildesq
    \ee
}
use~\re{hsqfp} to replace $\htildesq$ by 
$\frac{G_{1}(X)}{\left(1-\nicefrac{\omega}{6}\right)}$.

The eigenvalues ($\lambda_{i}$'s) of the matrix are:
\begin{equation}
\begin{aligned}
\label{eigenvalues}
\lambda_{i}&=-4, -4, -3,\\
&\nicefrac{1}{2} \left((M_{22}+M_{33})\mp \sqrt{({M_{22}-{M_{33}}})^2+4 {M_{23}} {M_{32}}}\right),\\
&i=1,\dots,5
\end{aligned}
\end{equation}
respectively. Using~\re{7th} we find that
\be
\label{m22m33}
M_{22}+M_{33}=-3\, .
\ee
We can thus impose the following conditions on elements~\re{M22}-\re{M33} so that the fixed-points become stable or stable spirals:
\begin{itemize}
    \item Stable solutions
    \be
    0\leq(3+2M_{22})^2+4M_{23}M_{32}<9\, ,
    \ee
    \item Stable spiral solutions
    \be
     (3+2M_{22})^2+4M_{23}M_{32}<0\, .
     \ee
\end{itemize}
These are equations~\re{stable} and~\re{stablespiral} in Section~\ref{stability}.
\section{Non-Standard Stability Analysis for Case 1}
\label{ModifiedStability}

In this appendix we provide the detailed computations supporting the linear stability analysis presented in Section \ref{case1 stability} and the numerical analysis explained in Section \ref{numCase1} for the Case 1 fixed points. 

In order to compute $\mathbf{\delta {V}}$ from \re{perturbed} and the limits \re{dNdeltaV} and \re{ybyg2}, we split the analysis in two steps:
\begin{itemize}
    \item[1.] we provide the analytical expression for the matrix ${\rm\bf P}$ and its inverse and analyze the fixed-point limit of their elements;
    \item[2.] we expand \re{perturbed} and show that, under certain conditions, perturbations are infinitesimally close to zero in the limit $N\to+\infty$ and $\mathbf{V}_{0}\to\mathbf{V}_{FP}$. 
    We then show that in linear perturbation theory the limits \re{dNdeltaV} and \re{ybyg2} cannot be assessed.
 \end{itemize}
 
 The third subsection of this appendix deals with the initial conditions needed in numerical analysis presented in Section \ref{scanning}.

\subsection{Matrix ${\rm\bf P}$ and its inverse in the fixed-point limit}

In order to analyze the fixed-point limit of $\mathbf{\delta {V}}$ we previously need to study the matrix ${\rm\bf P}$ and its inverse.

The column vectors in the matrix ${\rm\bf P}$ are composed of eigenvectors of ${\rm\bf M}$. The matrix ${\rm\bf P}$ reads
\[
{\rm\bf P}=
  \begin{bmatrix}
    P_{11} & P_{12} & 0 & 0 & 0 \\
    P_{21} & P_{22} & P_{23} & P_{24} & P_{25} \\
    0 & 1 & P_{33} & 1 & 1 \\
    0 & 0 & 1 & 0 & 0 \\
    1 & 0 & 0 & 0 & 0
  \end{bmatrix}
\]
where
\begin{equation}
P_{11}=\frac{{M_{25}} {M_{32}}}{({M_{22}}+4) {M_{31}}},
\end{equation}
\begin{equation}
P_{12}=\frac{{M_{23}} {M_{32}}}{({M_{22}}+4) {M_{31}}}-\frac{4+M_{33}}{M_{31}},
\end{equation}
\begin{equation}
P_{21}=-\frac{{M_{25}}}{({M_{22}}+4)},
\end{equation}
\begin{equation}
P_{22}=-\frac{{M_{23}}}{({M_{22}}+4)},
\end{equation}
\begin{eqnarray}
P_{23}&=&-{M_{24}} ({M_{33}}+3)\\ \nonumber
&\times&\left({M_{22}} {M_{33}}+3 {M_{22}}-{M_{23}} {M_{32}}+3 {M_{33}}+9\right)^{-1}\, ,
\end{eqnarray}
\begin{eqnarray}
P_{24}&=&-2 {M_{23}}\\ \nonumber
&\times&\left(\sqrt{{\left(M_{22}-M_{33}\right)}^2+4 {M_{23}} {M_{32}}}+{M_{22}}-{M_{33}}\right)^{-1}\, ,
\end{eqnarray}
\begin{eqnarray}
P_{25}&=&2 {M_{23}}\\ \nonumber
&\times&\left(\sqrt{{\left(M_{22}-M_{33}\right)}^2+4 {M_{23}} {M_{32}}}-{M_{22}}+{M_{33}}\right)^{-1}\, ,
\end{eqnarray}
\begin{eqnarray}
P_{33}&=&{M_{24}} \\ \nonumber
&\times&\left(\left({M_{22}} {M_{33}}+3 {M_{22}} +3 {M_{33}}+9\right)M^{-1}_{32}-{M_{23}}\right)^{-1}\, .
\end{eqnarray}

We first need to address the fixed-point limit of the matrix ${\rm\bf M}$ elements. We recall that in Case 1, although both $y$ and $G_{2}(X)$ vanish, 
the ratio $y/G_{2}(X)$ is finite and is given by~\re{yg2}. 
Therefore, one can see that the elements of the matrix, $M_{31}$ and $M_{32}$ 
(\re{MM31} and~\re{MM32} respectively) 
are divergent in the fixed-point limit\footnote{
    We use the phrase ``fixed-point limit'' loosely in this appendix, 
    since, as we discuss in Section \ref{scanning}, the Case 1 fixed point
    cannot be approached dynamically.
}, since both of them scale as $(\sim 1/y)$. All the other elements of ${\rm\bf M}$ are finite.

The non-zero elements of ${\rm\bf P}$  which have the terms $M_{31}$ and $M_{32}$, either depend on $M_{31}^{-1}$, $M_{32}^{-1}$ (both approaching $0$) or contain the ratio $M_{32}/M_{31}$  
which tends to
\be
\label{m32m31}
\frac{M_{32}}{M_{31}}\to-G'_{2}(X)\left(\frac{y}{G_{2}(X)}\right)\, ,
\ee
at the fixed point. Therefore the elements of ${\rm\bf P}$ are convergent at the fixed point. 

We now show that some of the elements of ${\rm\bf P}^{-1}$ are divergent.  The matrix ${\rm\bf P}^{-1}$ reads
\[
{\rm\bf P}^{-1}=
  \begin{bmatrix}
    0 & 0 & 0 & 0 & 1 \\
    R_{21} & 0 & 0 & 0 & R_{25} \\
    0 & 0 & 0 & 1 & 0 \\
    R_{41} & R_{42} & R_{43} & R_{44} & R_{45} \\
    R_{51} & -R_{42} & R_{53} & R_{54} & R_{55}
  \end{bmatrix}\, ,
\]
where
\begin{equation}
R_{21}=\frac{-({M_{22}}+4)} {D_{1}}\, ,
\end{equation}
\begin{equation}
R_{25}=\frac{{M_{25}} {M_{32}}}{D_{1}}\, ,
\end{equation}
\begin{equation}
R_{41}=\frac{M_{31}\left(N_{-}\right)\left(8+N_{+}\right)}{4D_{1}D_{2}}\, ,
\end{equation}
\begin{equation}
R_{42}=\frac{-M_{32}}{D_2}\, ,
\end{equation}
\begin{equation}
R_{43}=\frac{N_{-}}{2D_{2}}\, ,
\end{equation}
\begin{equation}
R_{44}=\frac{-M_{24}M_{32}\left(6+N_{+}\right)}{2D_{2}D_{3}}\, ,
\end{equation}
\begin{eqnarray}
R_{51}&=&\frac{-M_{31}\left(-8+N_{-}-2M_{22}\right)}{4D_{2}D_{1}}\\ \nonumber
&\times&\left(N_{+}-2M_{22}\right)\, ,
\end{eqnarray}
\begin{equation}
R_{53}=\frac{N_{+}-2M_{22}}{2D_{2}}\, ,
\end{equation}
\begin{equation}
R_{54}=\frac{M_{24}M_{32}\left(6-N_{-}+2M_{22}\right)}{2D_{2}D_{3}}\, ,
\end{equation}
\begin{equation}
R_{55}=\frac{M_{25}M_{32}\left(8-N_{-}+2M_{22}\right)}{2D_{1}D_{2}}\, ,
\end{equation}
where
\begin{eqnarray}
D_{1}&=&M_{31}}{{M_{22}} ({M_{33}}+4)\, ,\\ \nonumber
&-&{M_{23}} {M_{32}}+4 ({M_{33}}+4)\, ,
\end{eqnarray}
\begin{equation}
D_{2}=\sqrt{({M_{22}}-{M_{33}})^2+4 {M_{23}} {M_{32}}}\, ,
\end{equation}
\begin{eqnarray}
D_{3}&=&{M_{22}} ({M_{33}}+3)-{M_{23}} {M_{32}}\, ,\\ \nonumber
&+&3 {M_{33}}+9\, ,
\end{eqnarray}
\begin{eqnarray}
N_{-}&=&\sqrt{({M_{22}}-{M_{33}})^2+4 {M_{23}} {M_{32}}}\, ,\\ \nonumber
&+&{M_{22}}-{M_{33}}\, ,
\end{eqnarray}
\begin{eqnarray}
N_{+}&=&\sqrt{({M_{22}}-{M_{33}})^2+4 {M_{23}} {M_{32}}}\, \\ \nonumber
&+&{M_{22}}+{M_{33}}\, .
\end{eqnarray}
It is easy to see that $D_{1}$, $D_{3}$ $\sim \nicefrac{1}{y}$ and $D_{2}$, $N_{-}$,$N_{+}\sim \nicefrac{1}
{\sqrt{y}}$ which makes $R_{41}$, $R_{42}$, $R_{51}$ and $R_{52}$ divergent at the fixed point.

\subsection{Analysis of perturbations in the fixed-point limit}
The perturbations $\mathbf{\delta {V}}$ given by \re{perturbed} are expanded to
\begin{eqnarray}
\label{dely}
\delta{y}&=&\frac{f_{1}(\mathbf{V_0})}{4}\\ \nonumber
&+&\left(\frac{e^{-4N}}{4+M_{22}}\right)\times\biggl[M_{25}\frac{M_{32}}{M_{31}}C_{1}\\ \nonumber
&-&\biggl(\frac{16}{M_{31}}-M_{23}\frac{M_{32}}{M_{31}}\\ \nonumber
&+&4\frac{M_{33}}{M_{31}}+\frac{M_{22}}{M_{31}}(4+M_{33})\biggr)C_{2}\biggr]\, ,
\end{eqnarray}
\begin{eqnarray}
\label{delX}
\delta{X}&=&\frac{1}{12\left(M_{23-}M_{22}\frac{M_{33}}{M_{32}}\right)}\\ \nonumber
&\times&\biggl[-3M_{23}\left(\frac{M_{31}}{M_{32}}f_{1}(\mathbf{V_0})+\frac{4}{M_{32}}f_{3}(\mathbf{V_0})\right)\\ \nonumber
&+&\frac{M_{33}}{M_{32}}\left(12f_{2}(\mathbf{V_0})+4M_{24}f_{4}(\mathbf{V_0})+3M_{25}f_{5}(\mathbf{V_0})\right)\biggr]\\ \nonumber
&-&\left(\frac{e^{-4N}}{4+M_{22}}\right)(M_{25}C_{1}+M_{23}C_{2})\\ \nonumber
&-&M_{24}\left(\frac{3+M_{33}}{D_{3}}\right)C_{3}e^{-3N}\\ \nonumber
&-&\left(\frac{2M_{23}C_{4}e^{\lambda_{4}N}}{N_{-}}-\frac{2M_{23}C_{5}e^{\lambda_{5}N}}{N_{+}-2M_{22}}\right)\, ,
\end{eqnarray}
\begin{eqnarray}
\label{delDE}
\delta{\tilde{\Omega}_{DE}}&=&\frac{1}{12\left(M_{22}\frac{M_{33}}{M_{32}}-M_{23}\right)}\\ \nonumber
&\times&\biggl[-3M_{22}\left(\frac{M_{31}}{M_{32}}f_{1}(\mathbf{V_0})+\frac{4}{M_{32}}f_{3}(\mathbf{V_0})\right)\\ \nonumber
&+&\left(12f_{2}(\mathbf{V_0})+4M_{24}f_{4}(\mathbf{V_0})+3M_{25}f_{5}(\mathbf{V_0})\right)\biggr]\\ \nonumber
&+&C_{2}e^{-4N}+\frac{M_{24}C_{3}e^{-3N}}{D_{3}M_{32}^{-1}}\\ \nonumber
&+&\left(C_{4}e^{\lambda_{4}N}+C_{5}e^{\lambda_{5}N}\right)\, ,
\end{eqnarray}
\begin{equation}
\label{delm}
\delta{\tilde{\Omega}_{m}}=\frac{f_{4}(\mathbf{V_0})}{3}+C_{3}e^{-3N},
\end{equation}
\begin{equation}
\label{delr}
\delta{\tilde{\Omega}_{r}}=\frac{f_{5}(\mathbf{V_0})}{4}+C_{1}e^{-4N}.
\end{equation}
Since $\delta{X_{0}}$ and $\delta{\tilde{\Omega}_{DE_{0}}}$ are real quantities, we must set the imaginary terms in~\re{delX} and~\re{delDE} occurring in the last two terms to zero. From~\re{delDE}, this amounts to $C_{4}=C_{5}^{*}$. Additionally, it can be shown that this requirement also makes $\delta{X}$ real automatically.

We now study the limit of $\mathbf{\delta {V}}$ for $N\to+\infty$ and $V_{0}\to \mathbf{V}_{FP}$. We identify two different possibilities.
\begin{itemize}
    \item[(a)] If and only if the eigenvalues $\lambda_{4}$ and $\lambda_{5}$ are either real and negative or complex with negative real part, none of the two limits give rise to divergent terms. Thus the two limits commute and the final limit is well defined. Recalling that $f_{i}(\mathbf{V_{0}})$ vanish for $V_{0}\to \mathbf{V}_{FP}$, we have
\begin{eqnarray}
\lim_{\mathclap{\substack{N\to+\infty \\ V_{0}\to V_{FP}}}}\quad \mathbf{\delta V} \to 0\, .
\end{eqnarray}
We conclude that, under the above specified conditions for the eigenvalues, the perturbations approach zero near the fixed point.  
    \item[(b)] If the eigenvalues $\lambda_{4}$ and $\lambda_{5}$ are either real and positive (or complex with positive real part) the last term of \re{delDE} diverges (or is indeterminate) for $N \to \infty$ while the other terms remain finite. We conclude that $\delta{\tilde{\Omega}_{DE}}$ is divergent (or indeterminate).
 \end{itemize}

From a close inspection to $\lambda_{4}, \lambda_{5}$ given by~\re{eigenvalues} we find that, using~\re{yg2}, $M_{22}+M_{33}=-3$ (as in~\re{m22m33}). 
Thus, being ${M_{32}}$ divergent, condition (a) is realized only if ${M_{23}} {M_{32}} < 0$. We notice that
\begin{equation}
M_{23}=\frac{3X\left(1-\nicefrac{\omega}{6}\right)}{\omega \htildesq}\, ,
\end{equation} at the fixed point which is positive if $0<\omega<6$ and negative when $\omega>6$. This means one must satisfy the following condition for stable attractors:
\be
\label{m32ineq}
M_{32}\to
\begin{cases}
-\infty,\hskip 2pt 0<\omega<6\\
+\infty,\hskip 2pt \omega>6
\end{cases}\, .
\ee
To satisfy the above conditions at the fixed point, 
we first evaluate $M_{32}$ in~\re{MM32} at a point $X_{0}$ 
that is infinitesimally close to the fixed-point value of $X$. 
Demanding that 
\be
M_{32}
\begin{cases}
<0,\hskip 2pt 0<\omega<6\\
>0,\hskip 2pt \omega>6
\end{cases}
\ee
at the fixed-point limit, ~\re{m32ineq} will automatically be satisfied. 
Evaluating~\re{MM32} at $X_{0}$, 
we find that the divergent, hence dominating term, 
is $M_{32}^{\text{Div}}\equiv-M_{31}G_{2}{'}(X_{0})\left(\frac{y}{G_{2}(X_{0})}\right)$. 
As an illustration, 
consider the region $0<\omega<6$ and $\alpha_{\sigma}>0$. 
From~\re{ydef} and \re{defn}, we see that
\be
1-\left(\frac{y}{G_{2}(X)}\right)^{2}>0.
\ee
We can therefore write
\be
\text{Sign}\left(M_{31}\right)=\text{Sign}\left(\frac{G_{1}{'}(X_{0})}{y}\right)
\ee
Plugging this expression in the divergent term of $M_{32}$ gives 
\be
\text{Sign}\left(M_{32}^{\text{Div}}\right)=\text{Sign}\left(-\frac{G_{1}{'}(X_{0})G_{2}{'}(X_{0})}{G_{2}(X_{0})}\right)
\ee
which must be negative. 
We arrive at the relation~\re{reln1} in Section~\ref{case1 stability}
\be
G'_{1}(X_{0})G_{2}(X_{0})G'_{2}(X_{0})>0.
\ee
Similarly, we find that
\begin{itemize}
    \item $\alpha_{\sigma}>0$
    \begin{eqnarray}
    G'_{1}(X_{0})G_{2}(X_{0})G'_{2}(X_{0})&<&0,\hskip 2pt \omega>6\, ;
    \end{eqnarray}
    \item and $\alpha_{\sigma}<0$
    \begin{eqnarray}
    G'_{1}(X_{0})G_{2}(X_{0})G'_{2}(X_{0})&<&0,\hskip 2pt 0<\omega<6 \\
    G'_{1}(X_{0})G_{2}(X_{0})G'_{2}(X_{0})&>&0,\hskip 2pt \omega>6\, .
    \end{eqnarray}
\end{itemize}
which are~\re{reln2},~\re{reln3} and~\re{reln4} in Section~\ref{case1 stability}. Therefore, since $G_{2}(X)$ changes sign at $X_{FP}$, in a neighborhood of a fixed point $V_{FP}$, there are regions that satisfy (a) and regions that satisfy (b).

As underlined in Section \ref{case1 stability}, for $V_{FP}$ to be an attractor we require that
\begin{eqnarray}
\lim_{\mathclap{\substack{N\to+\infty \\ V_{0}\to V_{FP}}}}\quad \frac{d}{dN}\mathbf{\delta V} \to 0 \, .
\end{eqnarray}
When condition (a) is satisfied, the last term of the $d \delta{\tilde{\Omega}_{DE}}/dN$ (see \re{delDE}) is indeterminate in the limit. 

The last requirement for $V_{FP}$ to be an attractor is that, in linear perturbation theory, the following limit
\begin{eqnarray}
\label{yg2limit}
\lim_{\mathclap{\substack{N\to+\infty \\ V_{0}\to V_{FP}}}}\quad \left(\frac{y}{G_{2}(X)}\right)^{2}\, 
\end{eqnarray}
 approaches dynamically its fixed-point limit given by \re{yg2}. When condition (a) holds true we can expand \re{yg2limit} to
\begin{eqnarray}
\label{yg2PT}
\lim_{\mathclap{\substack{N\to+\infty \\ V_{0}\to V_{FP}}}}\quad \left(\frac{y_{0}+\delta{y}}{G_{2}(X_{0})+G'_{2}(X_{0})\delta{X}}\right)^{2}\, ,
\end{eqnarray}
and substituting \re{dely} and \re{delX} we find that the limit is indeterminate.

\subsection{Initial conditions for the numerical analysis}

In order to numerically investigate whether a solution of the dynamical equations is approaching the fixed point 
we need to set the initial conditions for the dynamical variables in a neighborhood close to the fixed-point value. In this regard
we perturb the dynamical variables around their fixed-point values 
instead of around their values \textit{close} to the fixed point, $\mathbf{V_{0}}$. 
This is possible because any point in the neighborhood of a solution 
infinitesimally close to the fixed point 
can be treated as a perturbation around the fixed-point value. 
We set the initial values 
$\left(X_{i},\tilde\Omega_{DE_{i}},\tilde\Omega_{m_{i}},\tilde\Omega_{m_{i}}\right)$ 
as follows:
\begin{equation}
X_{i}=X_{\pm}+\Delta X,
\end{equation}
\begin{equation}
\tilde\Omega_{DE_{i}}=\frac{G_{1}(X_{\pm})}{\left(1-\nicefrac{\omega}{6}\right)}+\Delta\tilde\Omega_{DE},
\end{equation}
\begin{equation}
\tilde\Omega_{m_{i}}=\Delta\tilde\Omega_{m},
\end{equation}
\begin{equation}
\tilde\Omega_{r_{i}}=\Delta\tilde\Omega_{r}
\end{equation}
We observe that, if we set $y_{i}\!=\!\Delta y$, then
the term $\frac{y}{G_{2}(X)}\equiv z(y,X)$ in~\re{rdef} 
would not be close to its fixed-point value given by~\re{yg2}. 
This happens because, on expanding $z(y,X)$ around the fixed point, we obtain
\begin{eqnarray}
z(\Delta y,X_{\pm}+\Delta X)&=&z(0,X_{\pm})\\ \nonumber
&+&\frac{\Delta y}{G_{2}(X_{\pm})}
-z(0,X_{\pm})\frac{G'_{2}(X_{\pm})}{G_{2}(X_{\pm})}\Delta X
\end{eqnarray}
The second and third terms in the above expansion 
have a divergent factor of $\frac{1}{G_{2}(X_{\pm})}$. 
Hence the initial value of $r$ obtained from~\re{rdef} 
is not close to its fixed-point value~\re{rfp}. 
We instead add a small perturbation to the fixed-point value of $r$, 
$r_{i}\!=\!\left(1+\frac{\omega G_{1}(X)}{XG'_{1}(X)\left(1-\nicefrac{\omega}{6}\right)}\right)
+\Delta r$ and use this value in~\re{ydef} to set the initial value of $y$ as
\be
    \label{yini}
    y_{i}= \pm\sqrt{1-\alpha_{\sigma}
        \frac{6\left(\tilde{\Omega}_{DE_{i}}-G_{1}(X_{i})\right)}
            {\omega r^{2}_{i} X_{i}^{2}}}G_{2}(X_{i})\, .
\ee

\section{Minkowski limit}
\label{Minkowski}

During evolution of the dynamical variables, 
if we encounter $H=0$, 
then the infinitesimal $dN$ is zero, 
making the standard evolution equations  indeterminate. 
In such a case, we revert to equations written in cosmic time. 
For notational convenience, we define ${\cal T}=tm_{g}$ 
and the following dimensionless quantities: 
$\tau=\left(\frac{{\sigma'}}{M_{Pl}}\right)^2$,
${\bar\rho}_m=\frac{\rho_m}{M_{Pl}^2 m_{g}^2}$,
${\bar\rho}_r=\frac{\rho_r}{M_{Pl}^2 m_{g}^2}$ 
and $\tilde{h}=\frac{{a'}}{a}$, 
where the primes ($'$) represent derivative with respect to ${\cal T}$. 
The resulting background evolution equations are:
\begin{equation}
\label{ydot}
{y'}=-4y \tilde{h},
\end{equation}
\begin{equation}
{X'}=X\left(\pm \sqrt{\tau}-\tilde{h}\right),
\label{xtau}
\end{equation}
\begin{eqnarray}
\label{tautau}
{\tau{'}}&=&-6\biggl[\tilde{h}\tau\\ \nonumber
&+&\frac{XG'_{1}(X)}{\omega}\left(\pm \sqrt{\tau}-r \tilde{h}\right)\biggr]\, ,
\end{eqnarray}
\begin{equation}
\label{mdot}
{\bar\rho}'_{m}=-3\tilde{h}{\bar\rho}_m,
\end{equation}
\begin{equation}
\label{rdot}
{\bar\rho}'_{r}=-4\tilde{h}{\bar\rho}_r.
\end{equation}
where one must substitute
\be
r=\sqrt{\frac{\tau\alpha_{\sigma}}{X^{2}\left(1-\left(\frac{y}{G_{2}(X)}\right)^2\right)}}\, ,
\label{rtau}
\ee
and
\be
\htildesq=\frac{\omega\tau}{6}+G_{1}(X)+\frac{{\bar\rho}_m}{3}+\frac{{\bar\rho}_r}{3}\,.
\label{hcosmic}
\ee
Notice that using ${\cal T}$ as the independent variable 
rather than $N$ provides us with a system of equations 
with no divergences as $\tilde{h}$ approaches zero, 
making it suitable to study the special fixed-point case $\tilde{h}=0$. 
However, these equations are inconvenient 
for the general stability analysis,
since the stability matrix has fewer diagonal terms. 

\subsection{Fixed point solutions}
We now focus on the $\tilde{h}=0$ fixed-points analysis. One can easily verify from~\re{ydot},~\re{mdot} and~\re{rdot}
that the derivatives of ${y}$, ${\bar{\rho}_{m}}$ and ${\bar{\rho}_{r}}$ 
approach zero.
These variables therefore approach constant values (and not necessarily $0$)
at the fixed point.

From~\re{xtau}, at least one of $X$ and $\tau$ must vanish.
There are three  cases:
\begin{itemize}
    \item If  $\tau=0$ and $X\neq0$, then~\re{tautau} 
    is automatically satisfied and from the Friedmann equation~\re{hcosmic} 
    \be
    \label{Friedmanh0t0}
    G_{1}(X)+\frac{{\bar\rho}_m}{3}+\frac{{\bar\rho}_r}{3}=0\,.
    \ee
    From~\re{defn}, we get $n(t)^{2}=(\dot{\phi}_{0})^{2}$. 
    From~\re{ydef}, we get $\left(\frac{y}{G_{2}(X)}\right)^{2}=1$, 
    which makes $r$ from~\re{rtau} indeterminate at the fixed point. 
    $r$ is also indeterminate from~\re{tautau}. 
    
    A special case occurs when $G_{1}(X)=0$. From~\re{Friedmanh0t0}, we get $\rho_{m}=\rho_{r}=0$. Since the initial values of matter and radiation density are not zero, we find that $a\to\infty$. 
    \item If $X=0$ but $\tau\neq0$, then~\re{tautau} is automatically satisfied 
    (meaning $\tau$ is indeterminate) and~\re{hcosmic} gives us
    \be
    \frac{\omega\tau}{6}+G_{1}(0)+\frac{{\bar\rho}_m}{3}+\frac{{\bar\rho}_r}{3}=0\,.
    \ee
    From~\re{xsigma}, 
   either $\sigma\to-\infty$ 
   (in which case, from~\re{defn} and~\re{rna}, $n$ and $r$ become divergent respectively),
   or
    $a\to\infty$ (in which case, from~\re{rna}, $r$ becomes divergent).
    \item If $\tau=X=0$,  then~\re{hcosmic} gives us
    \be
    G_{1}(0)+\frac{{\bar\rho}_m}{3}+\frac{{\bar\rho}_r}{3}=0\, .
    \ee
    From~\re{xsigma}, 
    either $\sigma\to-\infty$ 
    (in which case, from~\re{defn}, $\frac{\dot{\phi}_{0}}{n}$ is indeterminate),
    or $a\to\infty$ (in which case  $n^2=(\dot{\phi^{0}})^2$).
\end{itemize}

We then conclude that ${\Omega}_{DE}$ 
balances the sum of matter and radiation energy densities.

\subsection{Fixed-point linear stability}

To study the linear stability of the fixed points we follow~\re{ydot}-\re{rdot} defining $\mathbf{V}=\left[{y},{X},{\tau},{\bar{\rho}_m},{\bar{\rho}_r}\right]^{T}$ 
as the dynamical variables 
and $\mathbf{f(\mathbf{V})}$ as the RHS of these evolution equations. 
The stability matrix \textbf{M} at any time ${\cal T}$ takes the form
\[
M=
  \begin{bmatrix}
  \label{Mh0}
    M_{11} & M_{12} & M_{13} & M_{14} & M_{15} \\
    0 & M_{22} & M_{23} & M_{24} & M_{25} \\
    M_{31} & M_{32} & M_{33} & M_{34} & M_{35} \\
    0 & M_{42} & M_{43} & M_{44} & M_{45} \\
    0 & M_{52} & M_{53} & M_{54} & M_{55}
  \end{bmatrix}
\]
where
\begin{equation}
M_{11}=-4\tilde{h},
\end{equation}
\begin{equation}
M_{12}=\frac{-2yG_{1}{'}(X)}{\tilde{h}},
\end{equation}
\begin{equation}
M_{13}=\frac{-y\omega}{3\tilde{h}},
\end{equation}
\begin{equation}
M_{14}=\frac{-2y}{3\tilde{h}},
\end{equation}
\begin{equation}
M_{15}=M_{14},
\end{equation}
\begin{equation}
M_{22}=\pm\sqrt{\tau}-\tilde{h}-\frac{XG_{1}{'}(X)}{2\tilde{h}},
\end{equation}
\begin{equation}
M_{23}=\frac{X}{2}\left(\pm\frac{1}{\sqrt{\tau}}-\frac{\omega}{6\tilde{h}}\right),
\end{equation}
\begin{equation}
M_{24}=\frac{-X}{6\tilde{h}},
\end{equation}
\begin{equation}
M_{25}=M_{24},
\end{equation}
\begin{equation}
M_{31}=\frac{6y\tilde{h}\sqrt{\frac{\alpha_{\sigma}\tau}{\left(1-\left(\nicefrac{y}{G_{2}(X)}\right)^2\right)^{3}}}G_{1}{'}(X)}{\omega (G_{2}(X))^{2}},
\end{equation}
\begin{eqnarray}
M_{32}&=&6\frac{\sqrt{\tau}}{\omega}\Biggl[\frac{\left(G_{1}{'}(X)\right)^2}{2\tilde{h}}\sqrt{\frac{\alpha_{\sigma}}{\left(1-\left(\nicefrac{y}{G_{2}(X)}\right)^2\right)}}\\\nonumber
&-&G_{1}{'}(X)\Biggl(\pm1+\frac{\sqrt{\tau}\omega}{2\tilde{h}}\\ \nonumber
&+&\frac{y^{2}\tilde{h}G_{2}{'}(X)\left(\frac{\alpha_{\sigma}}{\left(1-\left(\nicefrac{y}{G_{2}(X)}\right)^2\right)}\right)^{\nicefrac{3}{2}}}{\alpha_{\sigma}\left(G_{2}(X)\right)^{3}}\Biggr)\\\nonumber
&-&\left(\pm X-\tilde{h}\sqrt{\frac{\alpha_{\sigma}}{\left(1-\left(\nicefrac{y}{G_{2}(X)}\right)^2\right)}}\right)G_{1}{''}(X)\Biggr],
\end{eqnarray}
\begin{eqnarray}
M_{33}&=&-6\Biggl(\tilde{h}\\\nonumber
&+&\frac{\left(\pm X-\tilde{h}\sqrt{\frac{\alpha_{\sigma}}{\left(1-\left(\nicefrac{y}{G_{2}(X)}\right)^2\right)}}\right)G_{1}{'}(X)}{2\sqrt{\tau}\omega}\Biggr)\\ \nonumber
&-&\frac{\tau\omega}{2\tilde{h}}+\sqrt{\frac{\alpha_{\sigma}\tau}{\left(1-\left(\nicefrac{y}{G_{2}(X)}\right)^2\right)}}\frac{G_{1}{'}(X)}{2\tilde{h}},
\end{eqnarray}
\begin{eqnarray}
M_{34}&=&\frac{-\tau}{\tilde{h}}\\ \nonumber
&+&\frac{G_{1}{'}(X)}{\omega \tilde{h}}\sqrt{\frac{\tau\alpha_{\sigma}}{\left(1-\left(\nicefrac{y}{G_{2}(X)}\right)^2\right)}},
\end{eqnarray}
\begin{equation}
M_{35}=M_{34},
\end{equation}
\begin{equation}
M_{42}=\frac{-3\bar{\rho}_{m}G_{1}{'}(X)}{2\tilde{h}},
\end{equation}
\begin{equation}
M_{43}=\frac{-\omega\bar{\rho}_{m}}{4\tilde{h}},
\end{equation}
\begin{equation}
M_{44}=\frac{-3}{\tilde{h}}\left(\htildesq+\frac{\bar{\rho}_{m}}{6}\right),
\end{equation}
\begin{equation}
M_{45}=\frac{-\bar{\rho}_{m}}{2\tilde{h}},
\end{equation}
\begin{equation}
M_{52}=\frac{-2\bar{\rho}_{r}G_{1}{'}(X)}{\tilde{h}},
\end{equation}
\begin{equation}
M_{53}=\frac{-\omega\bar{\rho}_{r}}{3\tilde{h}},
\end{equation}
\begin{equation}
M_{54}=\frac{-2\bar{\rho}_{r}}{3\tilde{h}},
\end{equation}
\begin{equation}
M_{55}=\frac{-4}{\tilde{h}}\left(\htildesq+\frac{\bar{\rho}_{r}}{6}\right).
\end{equation}
We cannot find the eigenvalues of \textbf{M} analytically,
because the characteristic polynomial is fifth order. Therefore we cannot apply the method described in Section \ref{case1 stability}. We instead study the characteristic polynomial in the fixed-point limit. 
We find that in both fixed-point cases,  $\tau=0$ and $X=0$, 
the polynomial has indeterminate coefficients
(due to the presence of ratios of powers of $\tilde{h}$, $\tau$ or $1-\left(\frac{y}{G_{2}(X)}\right)^2$),
or divergent coefficients 
(due to inverses of powers of $\tilde{h}$, $\tau$ or $1-\left(\frac{y}{G_{2}(X)}\right)^2$). 
This makes the eigenvalues and thus the stability of this special case indeterminate. Therefore, the stability of this fixed point can only be assessed by a numerical analysis. However,  we consider such analysis as out of the scope of our present work.

\section{Scalar perturbations}
\label{AppendPerturb}

Since there are certain disagreements  between our equations for the perturbations and those in the literature, for the sake of completeness, here we present the pertinent  equations.

Regarding scalar sector, the discrepancy we find can be noted in comparing the constraint equations obtained from the variation of the  action quadratic in the  perturbations, with respect to  $\Phi$ and $B$,  we obtain (in Fourier space): 
\begin{eqnarray}
0&=&6 H \dot{\psi}+\frac{2k^2}{a^2}\left( \psi+ \frac{k^2 E}{6}+a B H \right)-\frac{\omega \dot{\sigma}\dot{\delta\sigma}}{M_{\rm Pl}} \\
&-&\frac{\delta \chi  \rho _{,\chi }}{M_{\text{Pl}}}-\frac{\dot{\delta \chi } (p+\rho )}{\dot{\chi } c_s^2 M_{\text{Pl}}}+ \frac{3 \psi }{M_{\text{Pl}}^2}\left(3 H^2 M_{\text{Pl}}^2-\rho -\frac{\omega}{2} \dot{\sigma }^2 \right)\nonumber\\
&+&3 \psi  m_g^2 \left(J+\alpha _3 (X-1) (2 X-1)+(X-6) X+3\right)\nonumber\\
&+&\frac{3 \Phi }{M_{\text{Pl}}^2}\left(\frac{p+\rho (1+c_s^2)}{3 c_s^2}-3 H^2 M_{\text{Pl}}^2+\frac{\omega}{2}  \dot{\sigma }^2 \right)\nonumber\\
&+& \Phi (X-1) m_g^2 \left(J+\alpha _3 (X-1)^2-3 X+3\right)\nonumber\\
 &-&3X m_g^2 \left(J(X)+\alpha _3 (X-1) X-2 X\right)  \delta \sigma\,, \nonumber
 \label{eqphi} \end{eqnarray}
 
\begin{eqnarray}
0&=&\frac{3 \delta \chi  \left(r^2-1\right) (p+\rho )}{a \dot{\chi } \omega  M_{\text{Pl}}}-\frac{6\left(r^2-1\right)}{a \omega }  \label{eqB}\\
&\times &\left(H \Phi - \dot{\psi} -\frac{k^2}{6}\dot{E} \right)
+\frac{3}{\omega } B (r-1) m_g^2 \nonumber\\
&\times&\left[\alpha _3 (X-1) \left(r (X-1)^2-2 X+1\right)\right.\nonumber\\
&-&\left. 3 r (X-1)^2-X^2+6 X-3\right]\nonumber\\
&+&\frac{3 B \left(r^2-1\right) }{ \omega  M_{\text{Pl}}^2}\left(-3 H^2 M_{\text{Pl}}^2+ \rho +\frac{ \omega }{2}\dot{\sigma }^2\right)\nonumber\\
&+&\frac{3 }{\omega }B J (r-1) m_g^2 (r (X-1)-1)\nonumber\\
&+&\frac{3 \delta \sigma   \dot{\sigma }} {a X \omega  M_{\text{Pl}}}\Big{[}(r^2-1) X \omega-(r-1)\alpha _{\sigma } 
\nonumber\\ &&\times\left(J+\alpha _3 (X-1) X-2 X\right)\Big{]}\,.\nonumber
 \end{eqnarray} 
In deriving these equations we have not made use of the background equations.  One can show that using the assumed background evolution given in (28) and (29) of \cite{Gumrukcuoglu:2016hic}, these equations are equivalent to (48)  of \cite{Gumrukcuoglu:2016hic}.  Even after using the background equations, one can show that our eq.~(\ref{eqB}) differs from eq.~(4.22)  of   \cite{Heisenberg:2015voa} (we have an additional  contribution proportional to $\alpha_{\sigma}\delta\sigma$), and  (48)   of \cite{Gumrukcuoglu:2016hic} also differ also from those in \cite{Heisenberg:2015voa}  in their regime of applicability.  For this reason we believe our equations are correct.

\section{Stability of perturbations}
\label{perturbations}
In this appendix we provide the conditions obtained from requiring the stability of the perturbations, which   we used to restrict the values of the parameters of the theory in the numerical study of Section \ref{scanning}.

The mass of tensor perturbations is given by 
\begin{eqnarray}
\label{mgw}
m_{GW}^2&=&m_{g}^{2}X\big(3+3\alpha_{3}+\alpha_{4}+r X^{2}(\alpha_{3}+\alpha_{4})\\ \nonumber
&-&(1+r)X(1+2\alpha_{3}+\alpha_{4})\big).
\end{eqnarray}

To avoid tachyonic instabilities, we require $m_{GW}^2>-H^2$.   
For tensor perturbations, 
we do not reproduce Eq.~(4.7) of~\cite{Heisenberg:2015voa}, 
nor the underlying
equation of motion for the $\sigma$ field, Eq.~(3.13).
    Equation~\re{mgw} agrees with~\cite{AdF-SM} and~\cite{Heisenberg:2015voa} 
    at the Case 1 fixed point, but not elsewhere.

For vector perturbations, 
we confirm the expression (eq.~(4.17)) in~\cite{Heisenberg:2015voa},
for the  coefficient of the vector kinetic term, and 
re-write it as
\be
\label{vector}
\kappa_{V}^{2}=
\frac{2m_{g}^{2}a^{2}XG'_{1}(X)}{\left(k^2(1+r)+2m_{g}^{2}a^{2}XG'_{1}(X)\right)}\,.
\ee
For the perturbations to be stable, we must have $k_{V}^{2}>0$ 
and a vector mass-squared that limits any tachyonic instability 
$m_V^2=\frac{m_{GW}^{2}}{\kappa_{V}^{2}}>-H^{2}$. 
We see from~\re{vector} that if we impose 
\be
G'_{1}(X)>0,
\ee
the kinetic coefficient obeys $0<\kappa_{V}^{2}<1$.
If $m_{GW}^{2}>-H^{2}$, as imposed above,
then this also ensures tachyonic stability for vector perturbations.

For Case 1, after integrating out the BD mode, we obtain the 
 conditions for stability of the scalar perturbations. Our results for this case reproduce the ones in ~\cite{Gumrukcuoglu:2016hic}. 
At this fixed point,~\re{mgw} reduces to
\be
    m_{GW}^2= \frac{m_{g}^{2}(r-1)^2 X^3+\omega H^2(r(X+1)-2)}{(r-1)(X-1)}.
    \label{mgwg20}
\ee

Following~\cite{Gumrukcuoglu:2016hic}, we arrive at the stability conditions for scalar perturbations,
\be
    X^{2}<\alpha_{\sigma}\htildesq<r^{2} X^{2}\,.
    \label{scalarstabilityCase10ltomegalt6}
\ee
It follows that
\be
    r>1\,
    \label{scalarstabilityCase1rconstraint}
\ee
and 
\be
    \alpha_\sigma>0\,.
    \label{scalarstabilityCase1alphasigmaconstraint}
\ee

For simplicity,  we will  not evaluate the stability conditions for scalar perturbations in the region $\omega\geq\rm{6}$  
for neither $\alpha_{\sigma}>0$ nor $\alpha_{\sigma}<0$.

\bibliographystyle{hieeetr}
\bibliography{reference.bib}

\end{document}